\documentclass[a4paper,11pt]{article}
\usepackage{jinstpub} 

\usepackage{graphicx}
\usepackage[labelfont=bf, position=top]{subcaption}

\usepackage{amsmath}

\title{A novel perspective on crystal electromagnetic calorimeter design for the CEPC}

\author[a,b]{Weizheng Song,}
\author[a,b]{Yang Zhang,}
\author[a,b,c,1]{Shengsen Sun,\note{Corresponding author.}}
\author[a,e]{Fangyi Guo,}
\author[a,b]{Yuanzhan Wang,}
\author[a,c]{Linghui Wu,}
\author[d]{Jie Guo,}
\author[a,c]{Shaojing Hou,}
\author[a,c]{Yong Liu,}
\author[a,c]{Quan Ji,}
\author[a,c]{Jinfan Chang,}
\author[a,b,c]{Yifang Wang}
\affiliation[a]{Institute of High Energy Physics, Chinese Academy of Sciences, Beijing, 100049, China}
\affiliation[b]{University of Chinese Academy of Sciences, Beijing, 100049, China}
\affiliation[c]{High Energy Research Center, Henan Academy of Sciences, Zhengzhou, 450046, China}
\affiliation[d]{Luoyang Mining Machinery Engineering Design Institute CO.,LTD, Luoyang, 471000, China}
\affiliation[e]{China Center of Advanced Science and Technology, Beijing, 100190, China}

\emailAdd{sunss@ihep.ac.cn}

\abstract{
Crystal electromagnetic calorimeters (ECALs) are essential for high-precision measurements of electrons and photons in particle physics experiments. 
However, the conventional design, in which long crystal bars point radially toward the interaction region and lack longitudinal segmentation, is incompatible with the three-dimensional shower imaging required by Particle Flow Approach (PFA).
We propose a novel perspective on crystal ECAL design to address this limitation. 
The key innovation is a geometric reconfiguration in which crystal bars are oriented to face the interaction region and arranged orthogonally in adjacent longitudinal layers.
This layout achieves fine spatial segmentation of energy deposits by correlating measurements of orthogonal crystal bars.
An interleaved structure of regular and inverted trapezoidal modules is incorporated to maximize structural uniformity and detector hermeticity. 
This design is engineered to preserve the excellent intrinsic energy resolution of crystal ECALs while simultaneously providing the detailed three-dimensional shower imaging essential for PFA.
Simulation results confirm the feasibility of achieving excellent energy resolution of $1.14\%/\sqrt{E} \oplus 0.44\%$. 
Consequently, the proposed design repositions crystal ECAL as a foundational component for PFA-oriented detector systems at facilities such as the Circular Electron Positron Collider (CEPC), offering a new technical pathway to advance the physics goals of future colliders.
}

\keywords{crystal bar, ECAL, PFA, CEPC}

\begin{document}
\maketitle
\flushbottom
\section{Introduction}
\label{sec:intro}

With the discovery of the Higgs boson by the ATLAS and CMS collaborations at the Large Hadron Collider (LHC)~\cite{ATLAS:2012yve,CMS:2012qbp}, the final piece of the Standard Model (SM) has been put in place.
This milestone has shifted the focus of particle physics toward the precise testing of the SM and the search for new physics beyond it~\cite{deBlas:2944678,P5:2023wyd}.
To pursue these goals, several future electron‑positron collider projects have been proposed, such as the Circular Electron Positron Collider (CEPC)~\cite{CEPCStudyGroup:2023quu, GuimarãesdaCosta:2678417}, the Future Circular Collider (FCC‑ee)~\cite{FCC:2018evy}, the International Linear Collider (ILC)~\cite{Behnke:2013xla}, and the Compact Linear Collider (CLIC)~\cite{Aicheler:2012bya}.
To fully exploit the physics potential of the experiments at future high energy colliders, it is essential to explore complementary and diverse detector designs that can meet the stringent performance requirements.

The Particle Flow Approach (PFA)~\cite{Thomson:2007zza} is a promising strategy for achieving excellent jet energy resolution through the optimal combination of tracking and calorimetric information.
The ultimate jet energy resolution is constrained by the intrinsic performance of the detectors, and in particular by the ability to resolve overlapping energy deposits from closely spaced particles within dense jet cores.
These fundamental constraints therefore impose demanding requirements on electromagnetic calorimeters (ECALs) designed for PFA: excellent energy resolution, sufficient spatial segmentation for precise shower imaging, and compact longitudinal development for effective shower separation.
A high granularity sampling ECAL prototype based on silicon-tungsten (Si-W)~\cite{CALICE:2008kht, SiWECAL} has been developed by CALICE collaboration, providing three-dimensional shower imaging and precise spacial measurement, enabling excellent jet energy resolution combined with PandoraPFA~\cite{PandoraPFA}.
The high granularity calorimeter (HGCAL)~\cite{CMSHGCAL} has a sampling ECAL designed according to the PFA strategy and is a part of CMS upgrade program.
However, the energy resolution of sampling ECALs is intrinsically limited by fluctuations inherent to the sampling process.

Crystal ECALs have traditionally played an important role in precision measurements across a diverse range of particle physics experiments.
Their superior energy resolution and high detection efficiency are crucial in the successful physics programs of the experiments at both the intensity and high-energy frontiers~\cite{BESIII:2009fln,Belle-II:2010dht,CMS:2006myw}.
In conventional designs, long crystal bars are oriented radially toward the interaction region, providing fine transverse segmentation. 
Combined with a compact Moli$\grave{\mathrm{e}}$re radius ($R_M$), the crystal bar layout achieves excellent transverse position resolution.
This geometric design inherently lacks longitudinal segmentation, and consequently can not provide the detailed three-dimensional shower energy deposition information required by PFA.
A viable solution to this limitation is a crystal ECAL with fine three‑dimensional segmentation, realized through the use of small cubic crystals as the fundamental detection units~\cite{Dong:2017/C}.
However, achieving such granularity typically necessitates a large number of readout channels, which leads to increased power consumption, system complexity, and costs.
We propose a novel ECAL design based on long crystal bars as the fundamental detection units, offering a substantial reduction in readout channel count compared to the cubic crystal approach. 
The key innovation lies in a geometric reconfiguration that departs fundamentally from traditional radial-pointing layouts: crystal bars are oriented to face the interaction region and arranged orthogonally in adjacent longitudinal layers. 
The intersection points of orthogonal crystal bars in adjacent longitudinal layers define a three-dimensional grid of virtual cubes.
The energy deposited in each virtual cube is derived from the correlation of signals between orthogonal crystals and the shower profile, enabling three-dimensional shower imaging.
The continuous nature of the longitudinal energy profile, a direct consequence of electromagnetic and hadronic cascade dynamics, provides a powerful constraint for resolving ambiguities inherent to this orthogonal bar geometry. 
The proposed crystal-bar-based ECAL architecture enables access to three-dimensional shower energy deposition information while simultaneously preserving excellent energy resolution.
This design has consequently been adopted as the reference detector configuration for the CEPC~\cite{CEPCStudyGroup:2025kmw}.

The key properties of several commonly used inorganic scintillating crystals are summarized in Section \ref{Crystal ECAL}, bismuth germanate (BGO) emerges as the optimal trade-off candidate.
The crystal ECAL geometry, as presented in Section \ref{ECAL geometry design}, is designed to provide three-dimensional shower imaging and maximize structural uniformity and detector hermeticity.
This is achieved through an orthogonal arrangement of crystal bars in adjacent longitudinal layers and an interleaved configuration of regular and inverted trapezoidal modules.
Simulation and performance studies of the proposed ECAL geometry, presented in Section~\ref{simulation and performance}, confirm the capability to extract three‑dimensional shower energy depositions information while maintaining an electromagnetic (EM) energy resolution of $1.14\%/\sqrt{E} \oplus 0.44\%$.
In addition, the ambiguity removal method is also described.
Section~\ref{summary} summarizes the conclusions of this paper.

\section{Crystal ECAL}\label{Crystal ECAL}

Precise measurements of the Higgs boson properties and the electroweak observables at the CEPC place stringent requirements on the performance of the CEPC detector to identify and measure physics objects such as leptons, photons, jets and their flavors with high efficiency, purity and precision.
In particular, critical benchmark processes such as the $H\rightarrow\gamma\gamma$ decay~\cite{Guo:2022wti} and a comprehensive flavor physics program~\cite{Ai:2024nmn} demand excellent energy resolution of the ECAL system across a wide energy range.
Jets from Higgs, W, and Z boson decays need to be measured with a jet energy resolution of 3$\sim$5\%, allowing an average of $2\sigma$ or better separation of hadronic decays of these bosons, which necessitates some specific requirements on the material properties of the ECAL.
To achieve the required jet energy resolution, the PFA represents a promising strategy. 
Aiming at an unprecedented jet energy resolution, the momenta of charged particles in jets are measured in the tracking detectors with excellent resolution, energy deposits of both charged and neutral particles are obtained from the ECAL and hadronic calorimeter (HCAL) but typically have lower resolution compared to tracking detectors.
The limitation of the jet energy resolution arises from the imperfect association of energy deposits with the correct particles, rather than the performance of calorimeters themselves. 
A compact shower will undoubtedly improve the accuracy of matching charged tracks with showers, reduce misidentification arising from shower overlaps, and consequently enhance the jet energy resolution.
The material of the ECAL is demanded to have a short radiation length ($X_0$) and small $R_M$ which lead to compact EM showers.
It also has a large ratio of interaction length to $X_0$ ($\lambda_{I}$/$X_{0}$) which means that hadronic showers will tend to be longitudinally well separated from EM showers. 

Inorganic scintillating crystals are an ideal medium for the ECAL. 
Their high density and effective atomic number ensure compact EM shower development and efficient energy collection over a wide energy range. 
Combined with the logarithmic energy scaling of longitudinal shower development, this property supports a compact and cost‑effective detector geometry. 
Furthermore, the characteristically high light yield of inorganic crystals provides the photon statistics required to achieve excellent energy resolution. 
Beyond these optical‑material attributes, crystals offer favorable mechanical workability which facilitates the construction of finely segmented detector modules with complex geometries, including ease of cutting, polishing, and assembly.
Additional important advantages include intrinsic radiation hardness for stable performance in high luminosity environments, as well as fast scintillation decay times that enable precise timing measurements and pile‑up mitigation.
A comparison of key properties for several common inorganic crystals is presented in Table~\ref{tab:i}.

\begin{table}[htbp]
\centering
\caption{Properties of commonly used crystals.~\cite{CrystalCost}\label{tab:i}}
\smallskip
\begin{tabular}{c|cccc}
\hline
Crystals&CsI(Tl)&BGO&PWO&LYSO\\
\hline
Density $\rho$ (g/cm$^{3}$) & 4.51 & 7.13 & 8.3 & 7.4\\
Radiation Length $X_{0}$ (cm) & 1.86 & 1.12 & 0.89 & 1.14\\
Moli$\grave{\mathrm{e}}$re Radius $R_M$ (cm) & 3.57 & 2.23 & 2.00 & 2.07\\
Nuclear Interaction Length $\lambda_{I}$ (cm) & 39.3 & 22.7 & 20.7 & 20.9\\
$\lambda_{I}$/$X_{0}$ & 21.13 & 20.27 & 23.26 & 18.33\\
Light Yield (photons/MeV) & 58,000 & 7,400 & 130 & 30,000\\
Decay Time $\tau$ (ns) & 1220 & 300 & 30/10 & 40\\
Estimated Cost (\$/cc) & 4.0 & 6.0 & 7.5 & 32.0\\
\hline
\end{tabular}
\end{table}

Among the candidate inorganic crystals, BGO presents a well‑balanced set of properties for the CEPC ECAL. 
Compared to thallium‑doped cesium iodide (CsI(Tl)), BGO has a significantly higher density and shorter $X_0$ and $R_M$, resulting in more compact EM shower development. 
This intrinsic compactness permits a shallower calorimeter design, which in turn reduces the radial footprint of the ECAL, lowering the material and cost burden on downstream subsystems such as the HCAL and the magnet yoke. 
In contrast to lead tungstate (PWO), BGO provides a substantially higher light yield, directly improving the stochastic term of the energy resolution. 
Furthermore, while lutetium‑based crystals such as LYSO offer excellent performance, BGO maintains a considerably more favorable cost‑to‑performance ratio for large‑scale detector production. 
Taken together, these attributes position BGO as a technically sound and economically viable candidate that effectively reconciles the demands for high resolution, compact geometry, and project‑scale affordability at the CEPC.

\section{ECAL geometry design}\label{ECAL geometry design}

\subsection{ECAL layout} 

\begin{figure}[!htbp]
\centering
\includegraphics[width=.48\textwidth]{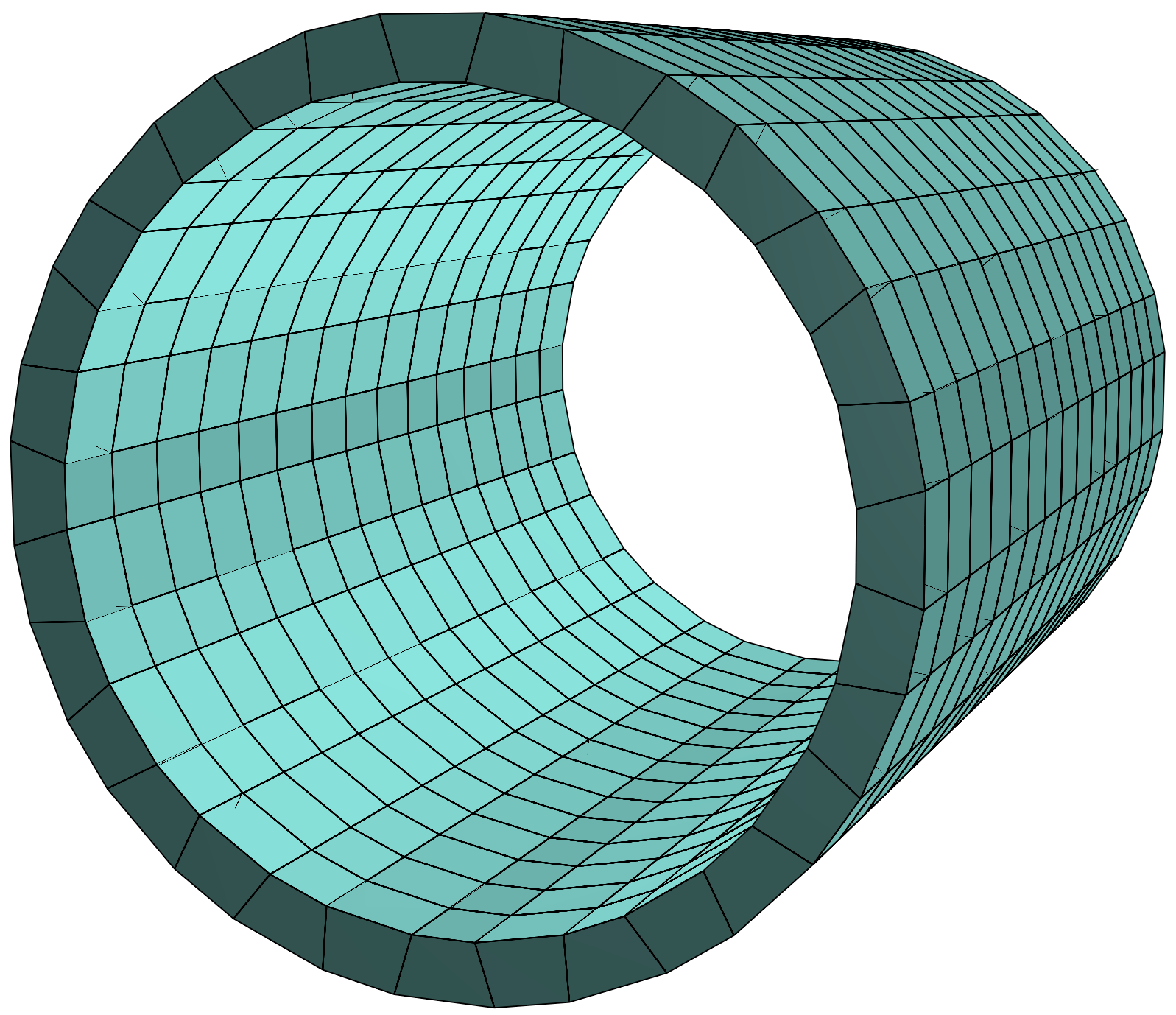}
\caption{The schematics of the barrel ECAL.}
\label{fig:Geometry}
\end{figure}

\begin{figure}[!htbp]
\centering
\includegraphics[width=.66\textwidth]{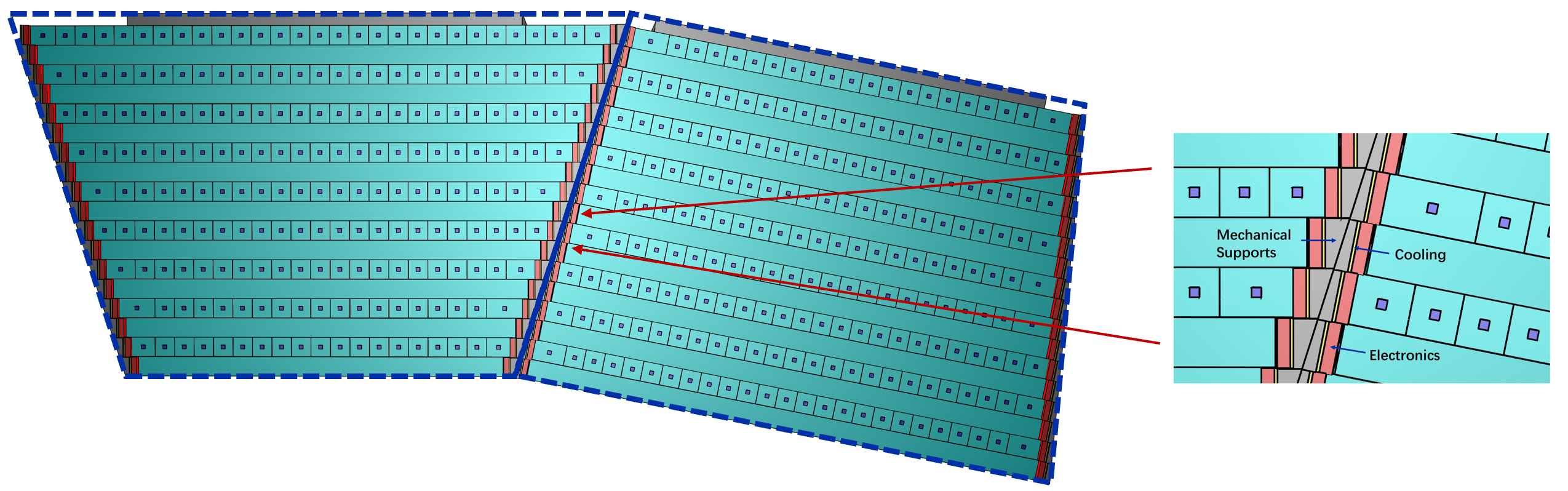}
\caption{Schematic diagram of the cross-sections of regular and inverted trapezoidal modules.}
\label{fig:twoTrapezoidal}
\end{figure}

As illustrated in the 3D schematic Figure \ref{fig:Geometry}, the barrel ECAL is constructed as a 32-sided polygonal prism with an inner radius of 1830 mm. 
Its radial depth of 300 mm accommodates the crystals, back-end electronics, and integrated cooling systems, extending to an outer radius of 2130 mm. 
With a length of 5800 mm along the beamline (\textit{Z}-axis), the detector provides a polar angular coverage of $|\cos\theta| < 0.85$.
The barrel ECAL comprises 480 modules, organized into 15 rings along the \textit{Z}-axis. 
Within each ring, the 32 modules are arranged azimuthally in the transverse (X-Y) plane.
The modules feature a trapezoidal cross-section in the transverse plane and are positioned in an alternating pattern of regular and inverted orientations as shown in Figure \ref{fig:twoTrapezoidal}. 
This design enhances the detector's hermeticity, reducing the probability of photon loss in inter-module regions that lack sensitive volume.
The interfaces between the modules are tilted by ${\pm}12^{\circ}$ relative to the radial direction with the sign determined by the orientations of the neighboring trapezoids.
The symmetry axes of both the regular and inverted trapezoids are aligned along the radial direction, pointing toward the interaction point (IP).
Only the long base of the inverted trapezoid is tangent to the circle of radius 1830 mm, and the short base of the regular trapezoid is tangent to the circle of radius 1835.9 mm.
The heights of the modules provide a thickness of around $24X_0$.

\subsection{ECAL module}

\begin{figure}[h]
\centering
\includegraphics[width=.75\textwidth]{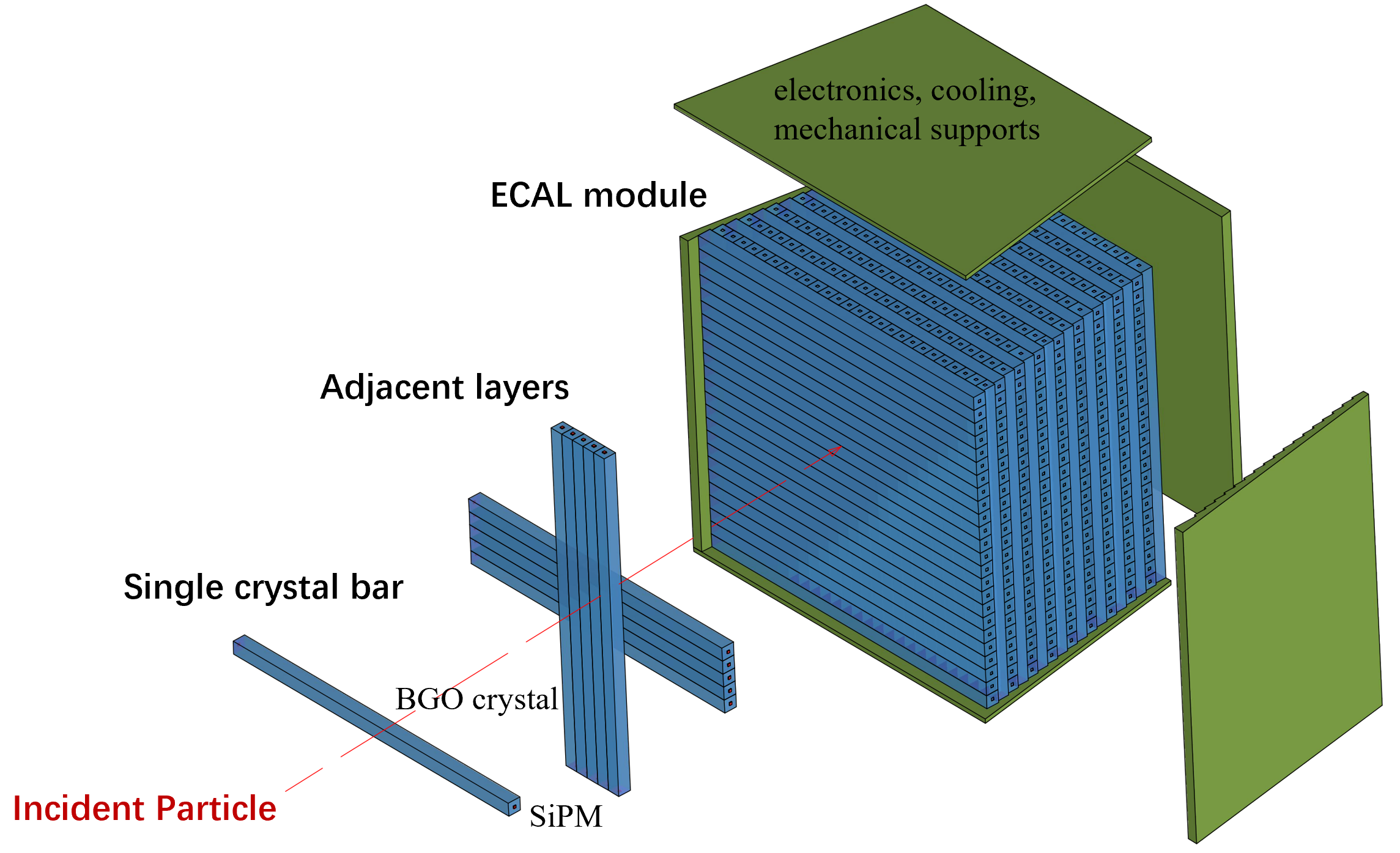}
\caption{The schematics of an ECAL module.}
\label{fig:bar2layer2module}
\end{figure}

Each ECAL module is radially segmented into 18 layers, with each layer consisting of an array of crystal bars positioned side-by-side.
The two arrays of crystal bars in adjacent layers are arranged to be orthogonal to each other, as shown in Figure \ref{fig:bar2layer2module}.
The crystals in the same layers of adjacent modules are precisely aligned at their interfaces to avoid introducing unnecessary complications for data processing.
Within each ECAL module, the array of crystals is surrounded by several service components including electronics, cooling and mechanical supports.
The front-end application-specific integrated circuits (ASICs) embedded in the printed circuit boards (PCBs) convert the charge and time stamp from the optoelectronic devices into digital signals.
Copper cooling plates dissipate heat generated by the front-end electronics toward the rear part of ECAL module.
Mechanical supports made of carbon fiber reinforced polymer (CFRP) enable the integration of the detector components.

\subsection{Crystal bar}

A rectangular BGO crystal bar is the fundamental detection unit with a cross-section of approximately 15.0 mm $\times$ 15.0 mm and 15.6 mm × 15.6 mm for the regular and inverted trapezoidal modules, respectively.
Crystals are individually wrapped with Enhanced Specular Reflector (ESR) foil.
Each crystal is read out by silicon photomultipliers (SiPMs) at both ends.
This dimension of 15 mm corresponds to roughly 2/3 of a $R_{M}$ and 4/3 of a $X_0$. 
While a smaller cross-section will improve position resolution, it will also substantially increase the number of readout channels and complicate the mass production of the crystal bars. 
The chosen size therefore represents a balance between position resolution, channel count, and manufacturing feasibility.

\begin{figure}[ht]
\centering
\begin{subfigure}[t]{0.48\textwidth}
    \includegraphics[width=\textwidth]{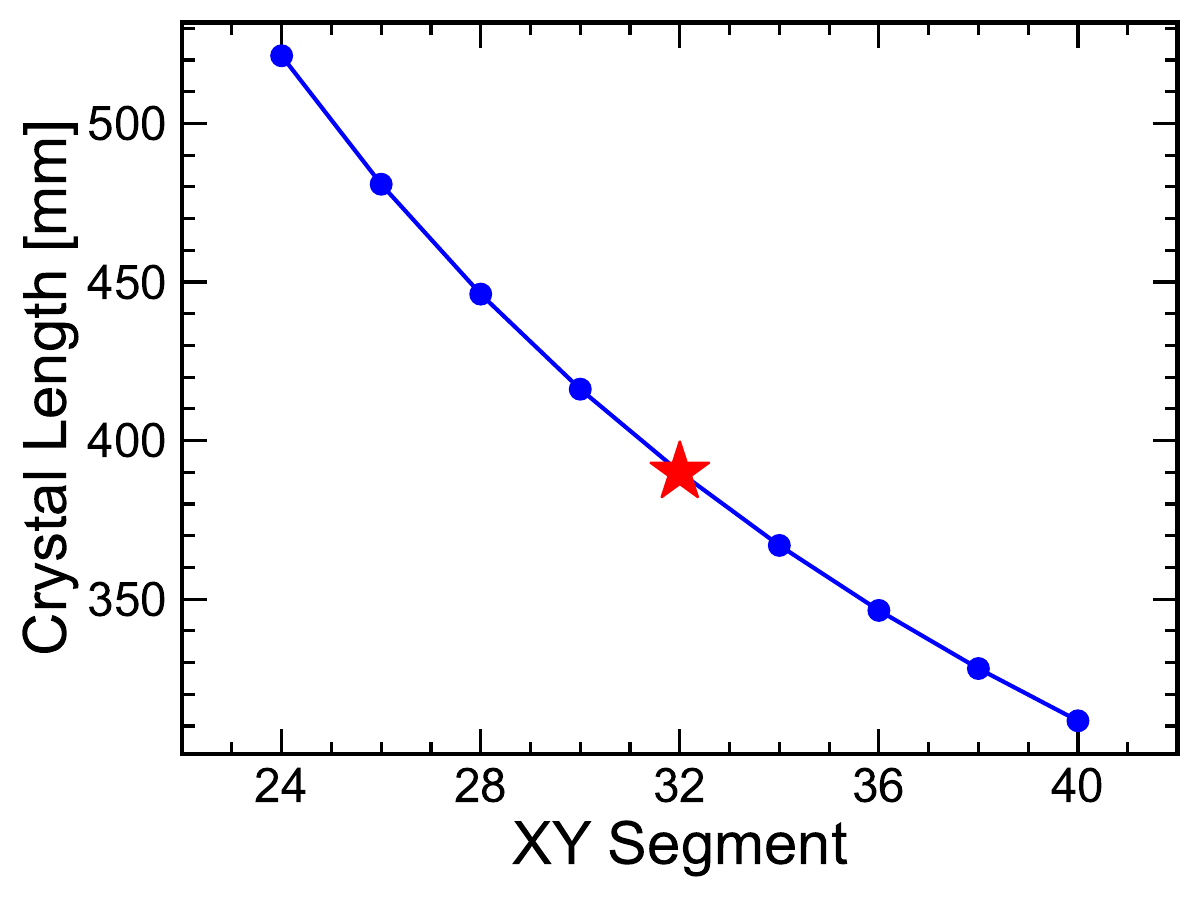}
    \caption{}
    \label{fig:xyBarLength}
\end{subfigure}
\hfill
\begin{subfigure}[t]{0.48\textwidth}
    \includegraphics[width=\textwidth]{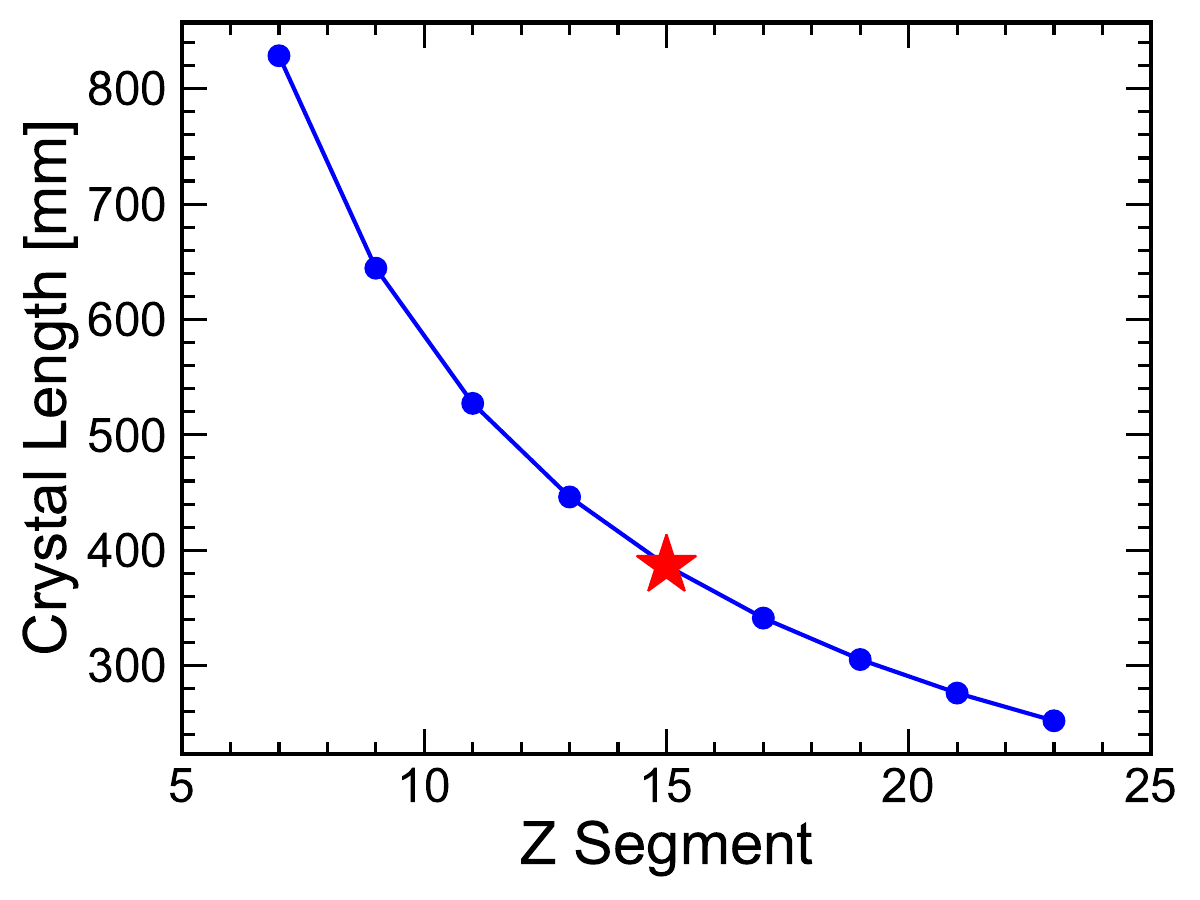}
    \caption{}
\label{fig:zBarLength}
\end{subfigure}
\caption{The corresponding lengths of long crystal bars for different numbers of barrel ECAL segments in XY and Z directions.}
\label{fig:BarLength}
\end{figure}

The lengths of the crystal bars are determined by the granularity of the barrel ECAL's segmentation in $\phi$ and $Z$, a design choice that embodies the compromise between performance objectives, manufacturing complexity, and overall cost.
Insufficient granularity in the transverse plane leads to an inefficient use of the space, thereby increasing the radial extent of the outer detector systems and the overall project cost.
Conversely, excessive segmentation results in an increased number of inter-module gaps and substantially raises the channel count of the readout electronics.
This trade-off is quantified in Figure \ref{fig:xyBarLength}, which shows the crystal bar length in the radial mid-layer (layer 9 of 18) as a function of the number of $\phi$ segments. 
Only an even number of $\phi$ segments are considered to preserve azimuthal symmetry.
A transverse segmentation of 32 is selected, resulting in a crystal bar length of 390 mm in layer 9.
A parallel constraint exists in the $Z$-direction. 
An odd number of $Z$-segments is mandatory to prevent a non-instrumented crack at the central detector plane ($z=0$). 
The dependence of the crystal bar length in the even-numbered layers on the $Z$-segmentation is shown in Figure \ref{fig:zBarLength}. 
To ensure the crystal bar lengths in the even-numbered layers remain comparable to that of layer 9, a value of 40 mm was selected, which is achieved with a Z-segmentation of 15.

\subsection{Crystal thickness}

\begin{figure}[ht]
\centering
\begin{subfigure}[t]{0.48\textwidth}
    \includegraphics[width=\textwidth]{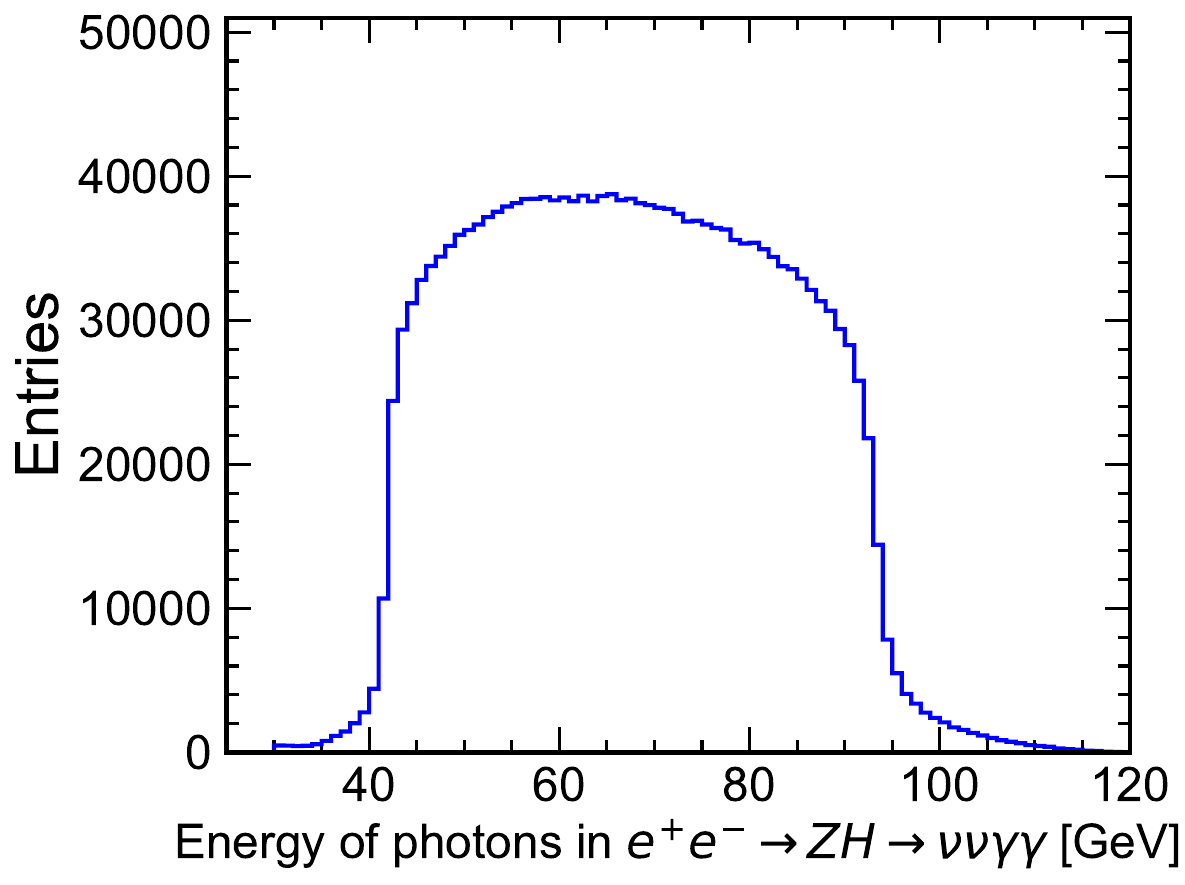}
    \caption{}
    \label{fig:EnergyPhotonsHaa}
\end{subfigure}
\hfill
\begin{subfigure}[t]{0.48\textwidth}
    \includegraphics[width=\textwidth]{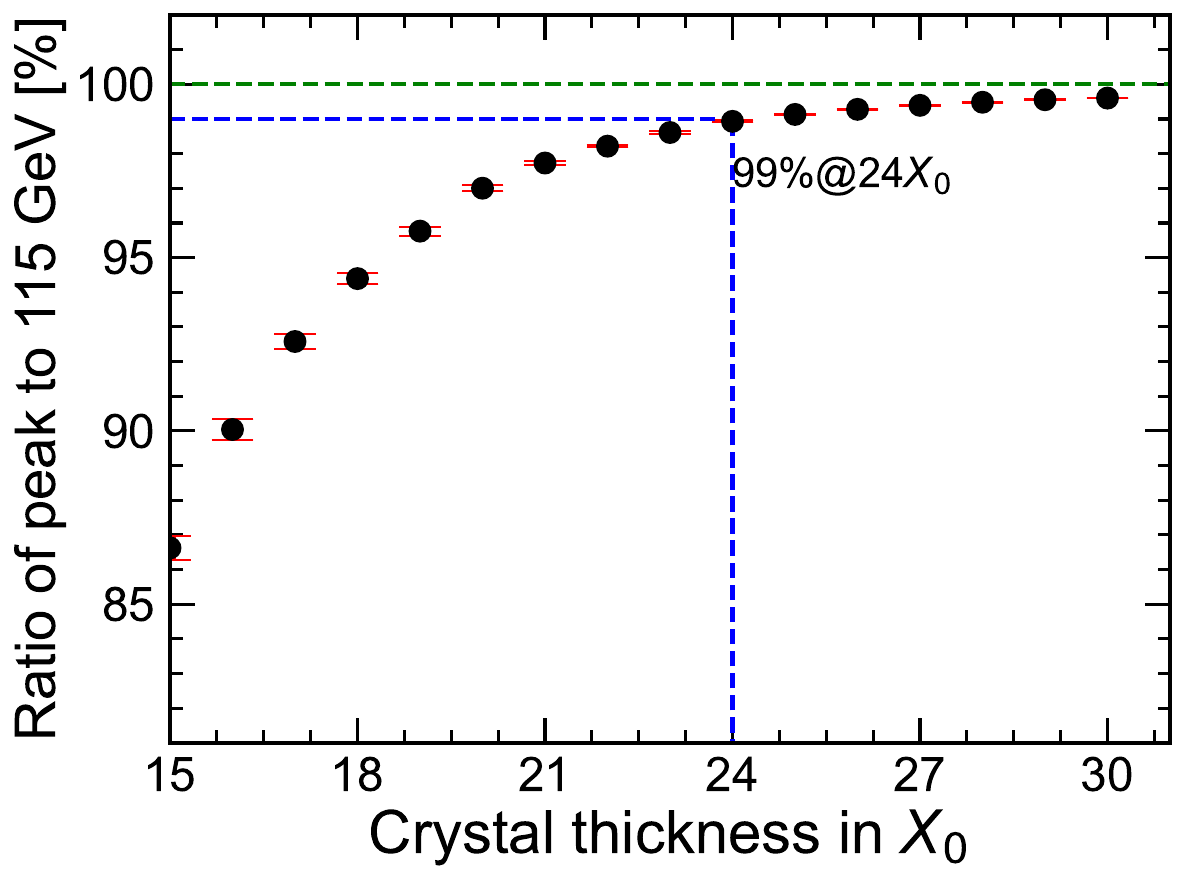}
    \caption{}
\label{fig:profile}
\end{subfigure}
\caption{(a) The energy distribution of the final-state photons in the physical process $e^+e^- \rightarrow ZH \rightarrow \nu \nu \gamma \gamma$ at the CEPC with 240 GeV center-of-mass energy; (b) The dependence of the ratio of peak deposited to the incident 115 GeV energy on the crystal thickness in $X_0$.}
\label{fig:Longitudinal depth}
\end{figure}

The cumulative thickness of the crystals is constrained by the requirement to achieve a specific energy resolution for incident high-energy particles.
Insufficient thickness leads to longitudinal leakage of the shower development, consequently resulting in failure to achieve the desired resolution.
This requirement is exemplified by the benchmark channel $e^+e^-{\rightarrow}ZH$ with $H\rightarrow\gamma\gamma$, whose precise measurement demands a high photon energy resolution central to the physics program.
Simulation of this process yields a maximum energy for the final-state photons of approximately 115 GeV, as shown in the energy deposition distribution in Figure~\ref{fig:EnergyPhotonsHaa}.
This energy also represents the approximate upper limit for final-state photons in other key physics benchmarks across the CEPC program.
The choice of crystal thickness is performed based on a series of simulations of interaction of 115 GeV photons within BGO crystals.
The dependence of the ratio of peak deposited to the incident 115 GeV energy on the crystal thickness in $X_0$ is shown as Figure~\ref{fig:profile}.
This result demonstrates that a crystal thickness of 24$X_{0}$ is sufficient to contain 99\% of the energy from the 115 GeV photons.
The total thickness of crystals of regular and inverted trapezoids are selected as 280.8 mm and 270.0 mm, corresponding to 25.1 $X_{0}$ and 24.1 $X_{0}$, respectively.

\subsection{Tilt angle}

\begin{figure}[ht]
\centering
    \includegraphics[width=0.75\textwidth]{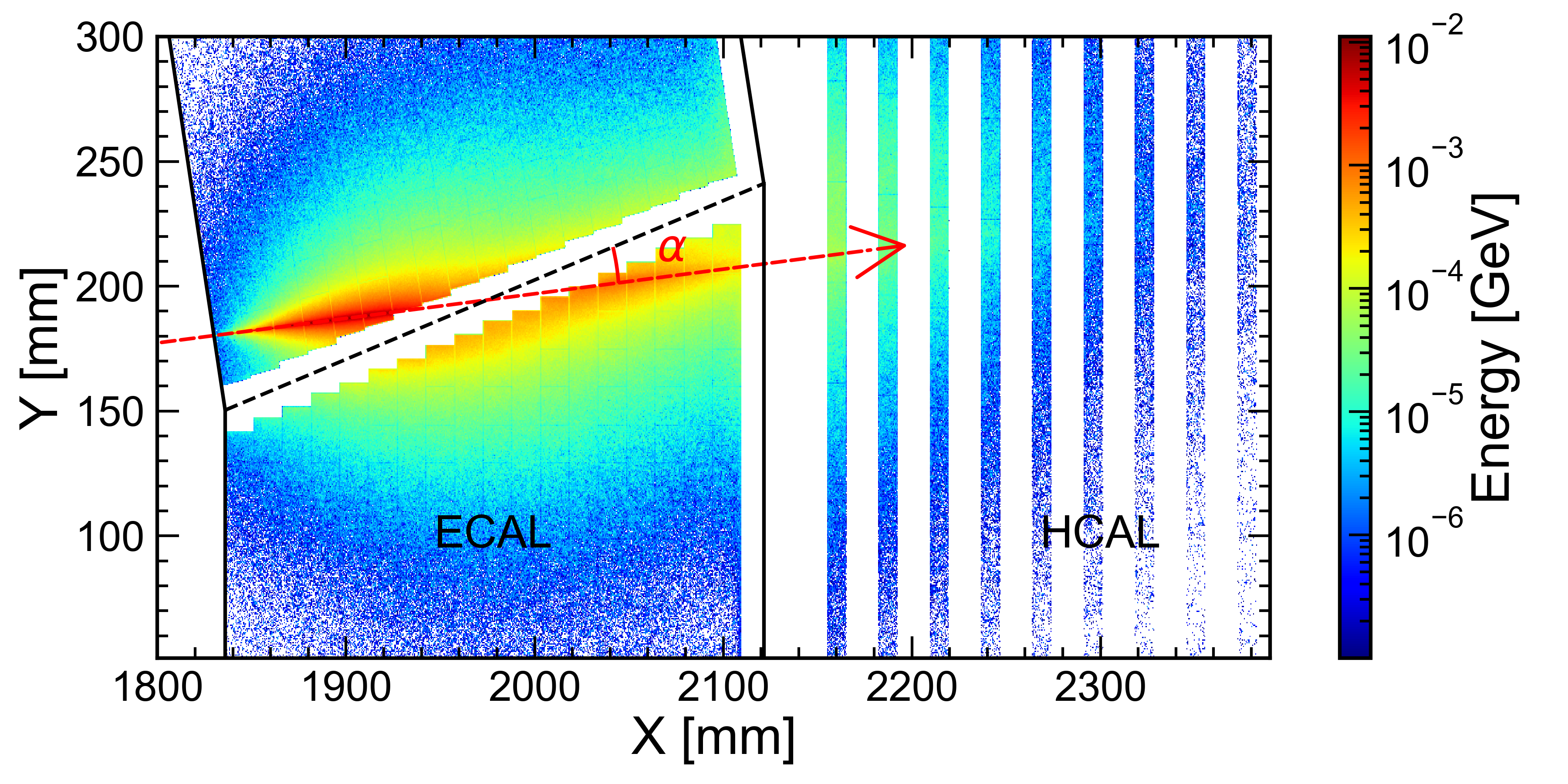}
\caption{Energy deposition of 20 GeV single photons in the Geant4 simulation, illustrating energy leakage due to cracks between the modules.}
\label{fig:EffiencyLoss}
\end{figure}

\begin{figure}[h]
\centering
\begin{subfigure}[t]{0.48\textwidth}
\includegraphics[width=\textwidth]{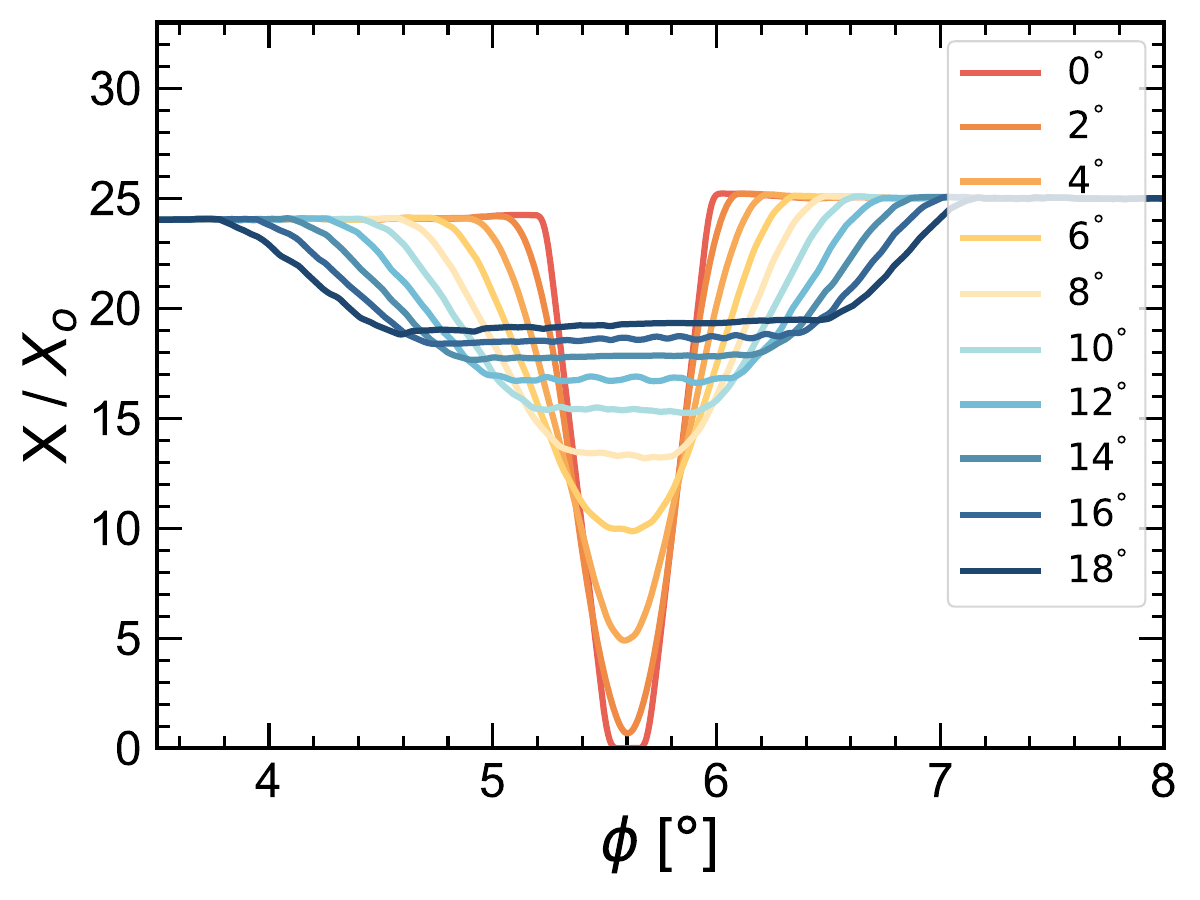}
    \caption{}
    \label{fig:X0Phi}
\end{subfigure}
\hfill
\begin{subfigure}[t]{0.48\textwidth}
\includegraphics[width=\textwidth]{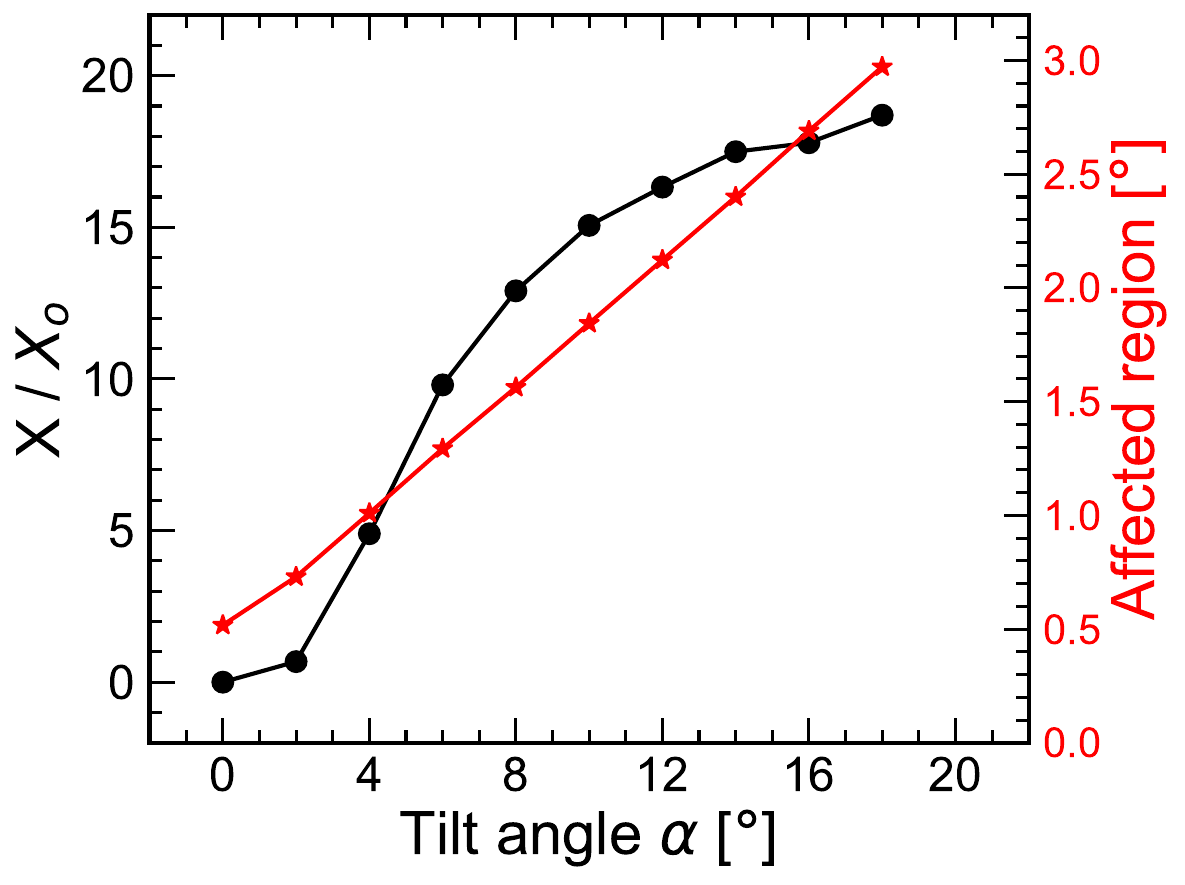}
    \caption{}
\label{fig:DegreeOptimal}
\end{subfigure}
\caption{(a) Dependence of the effective crystal thickness on the azimuthal angle; (b) Variations in effective crystal thickness and affected region as functions of tilt angle, with the thickness measured along the radial direction passing through the midpoint of the inter-module interface.}
\label{fig:degreeOp}
\end{figure}

The cracks between the modules of the regular and inverted trapezoids are filled with readout electronics, mechanical supports, cooling systems, and air.
The presence of passive materials in the inter-module cracks reduces the effective sensitive material thickness, and the resulting energy leakage degrades the performance of the ECAL.
Figure \ref{fig:EffiencyLoss} demonstrates the spatial distribution of energy deposition in the transverse plane from a simulated 20 GeV photon using Geant4~\cite{GEANT4:2002zbu}.
The regions to the left and right of the x-coordinate at 2130 mm correspond to the ECAL and the HCAL, respectively.
In the depicted ECAL section, the solid black lines indicate the inner and outer boundaries of the modules, and the white region between two adjacent modules represents the geometric crack of passive materials, the dashed black line denotes the inter-module interface.
The red dashed line indicates the direction of the incident photon, which originates from the IP and points toward the midpoint of the inter-module interface.
The tilt angle $\alpha$ is defined as the angle between the inter-module interface and the reference radial direction.
This reference radial direction passes through the interface midpoint and is identical to the photon direction depicted here.
The fact that the effective BGO crystal thickness in a certain radial direction remains constant for a given tilt angle establishes it as a robust evaluation metric.
Figure~\ref{fig:X0Phi} shows the dependence of the effective crystal thickness in units of $X_0$ on the azimuthal angle $\phi$ in transverse plane, where distinct colors represent different tilt angles.
The variations in both effective crystal thickness $X$ and affected region as functions of tilt angle are shown in Figure~\ref{fig:DegreeOptimal}, with the crystal thickness measured along the radial direction passing through the midpoint of the inter-module interface.
An increased tilt angle enhances crystal thickness, thereby reducing energy leakage, at the expense of enlarging the affected region.
A tilt angle of $12^{\circ}$ is therefore designated as the operating point that balances this trade-off.
As a preliminary measure, an energy correction algorithm~\cite{CyberPFA} has been introduced into the data processing chain to account for energy loss caused by particles traversing inter-module cracks.

\section{Simulation and performance}\label{simulation and performance}

\subsection{Detector description and digitization}
The barrel ECAL description is implemented within the DD4hep framework~\cite{Frank:2014zya} and follows the detector design demonstrated in Section~\ref{ECAL geometry design}. 
The implementation is structured in two layers: a declarative compact detector description and a procedural detector constructor. 
The parameters of the ECAL, including material definitions and geometric dimensions, are stored in an XML-based compact description. 
This description is subsequently interpreted by a dedicated C++ detector constructor, which builds the corresponding detailed geometry model in memory by instantiating elements from the generic DD4hep detector description toolkit.
For physics simulation, the DD4hep geometry is automatically translated into a Geant4-compatible format via the DDG4 extension. 
Simulations are subsequently performed with Geant4 using the FTFP\_BERT physics list~\cite{ALLISON2016186} to model the energy deposition of incident particles within the implemented detector geometry.

A parameterized digitization model has been developed to convert Monte Carlo truth-level energy deposition into digitized energy measurements of each electronic readout channel. 
First, the deposited energy is converted into the number of photoelectrons detected by a SiPM coupled to one end of crystal bar. 
In this stage, the scintillation light yield, propagation and attenuation within the crystal, light collection efficiency and the photon detection efficiency (PDE) of the SiPM have been collectively normalized and parameterized using the Minimum Ionizing Particle (MIP) response (corresponding to 300 photoelectrons per MIP).
Subsequently, the charge of the SiPM output, which is determined by the number of detected photoelectrons and the SiPM pixel gain, is processed and digitized into Analog-to-Digital Converter (ADC) counts, following the characteristic response of the specific SiPM-readout ASIC design.
The parameters used in the digitization model were extracted from the beam test experiments~\cite{Qi:2025dvo}. 

\begin{figure}[htbp]
\centering
\begin{subfigure}[t]{0.48\textwidth}
\includegraphics[width=\textwidth]{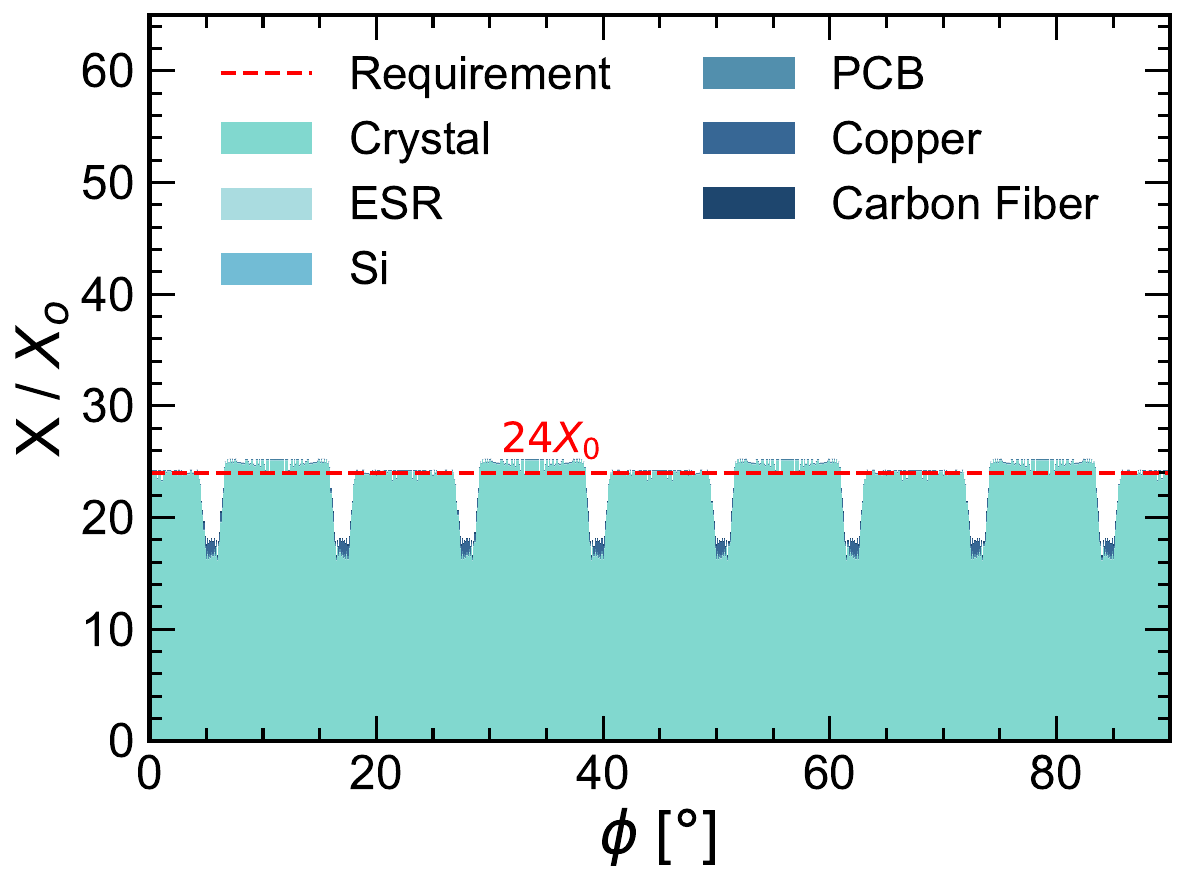}
    \caption{}
    \label{fig:PhiMaterialScan}
\end{subfigure}
\hfill
\begin{subfigure}[t]{0.48\textwidth}
\includegraphics[width=\textwidth]{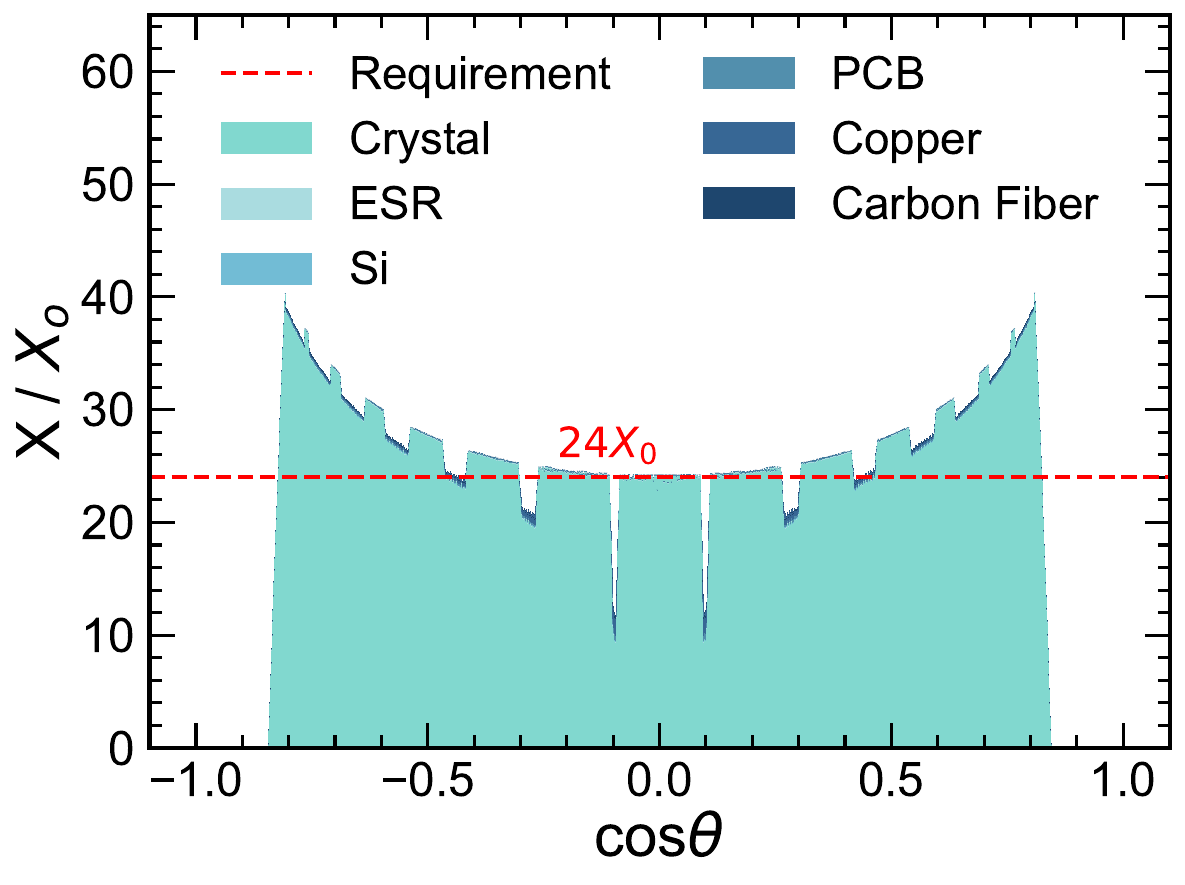}
    \caption{}
\label{fig:ThetaMaterialScan}
\end{subfigure}
\caption{Material budget scans of the barrel ECAL in the azimuthal ($\phi$) direction (a) and polar ($\theta$) direction (b).\label{fig:MaterialScan}}
\end{figure}

The spatial distributions of the material budget within the barrel ECAL are evaluated by conducting comprehensive scans across its full azimuthal ($\phi$) and polar ($\theta$) coverage, as shown in Figure~\ref{fig:MaterialScan}a and b.
The results demonstrate that the module arrangement effectively ensures the hermeticity of ECAL while maintaining the required thickness of approximately $24~X_0$ across the majority of the regions.
The interleaved structure of regular and inverted trapezoidal modules maximizes structural uniformity despite the unavoidable presence of inter-module cracks.

\subsection{Three-dimensional energy deposition information}

\begin{figure}[h]
\centering
\begin{subfigure}[t]{0.35\textwidth}
\includegraphics[width=\textwidth]{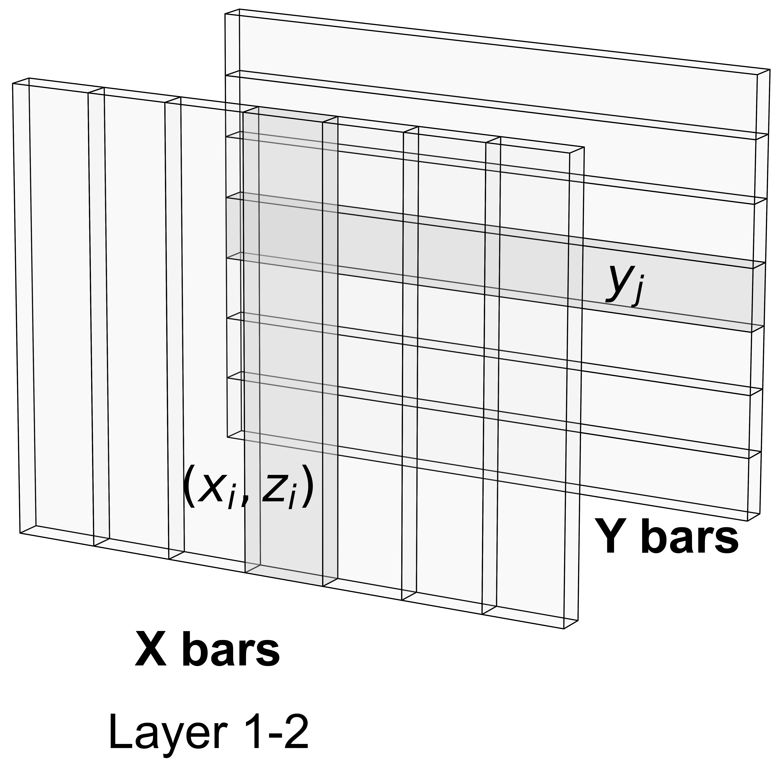}
    \caption{}
    \label{fig:segmentation_a}
\end{subfigure}
\begin{subfigure}[t]{0.35\textwidth}
\includegraphics[width=\textwidth]{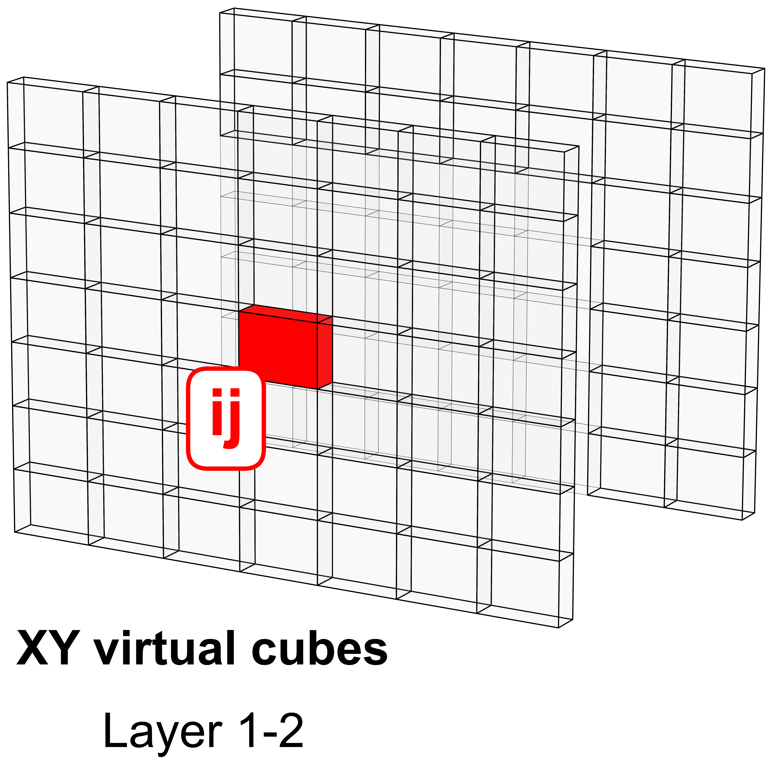}
    \caption{}
    \label{fig:segmentation_b}
\end{subfigure}
\hfill
\begin{subfigure}[t]{0.35\textwidth}
\includegraphics[width=\textwidth]{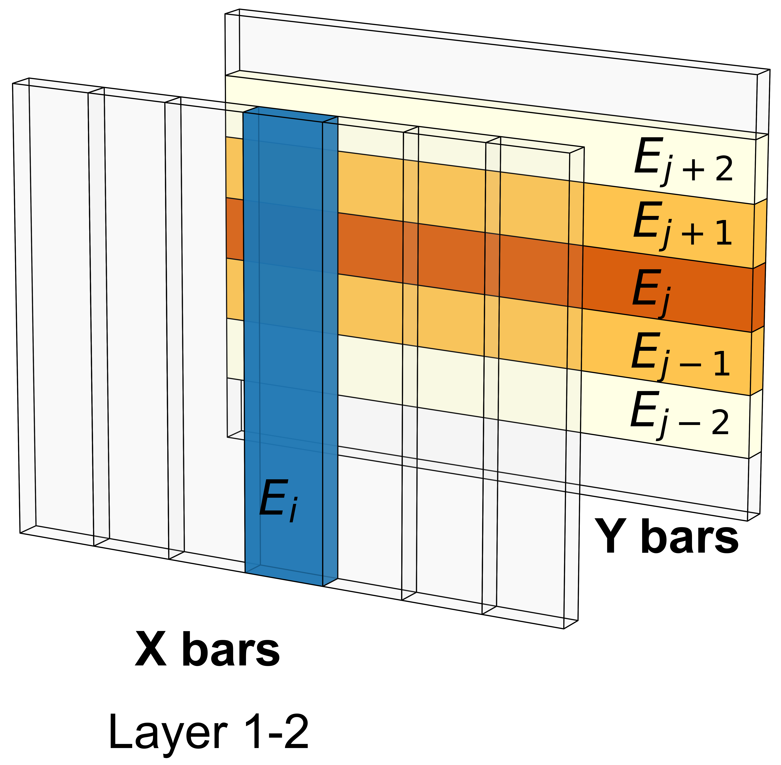}
    \caption{}
    \label{fig:segmentation_c}
\end{subfigure}
\begin{subfigure}[t]{0.35\textwidth}
\includegraphics[width=\textwidth]{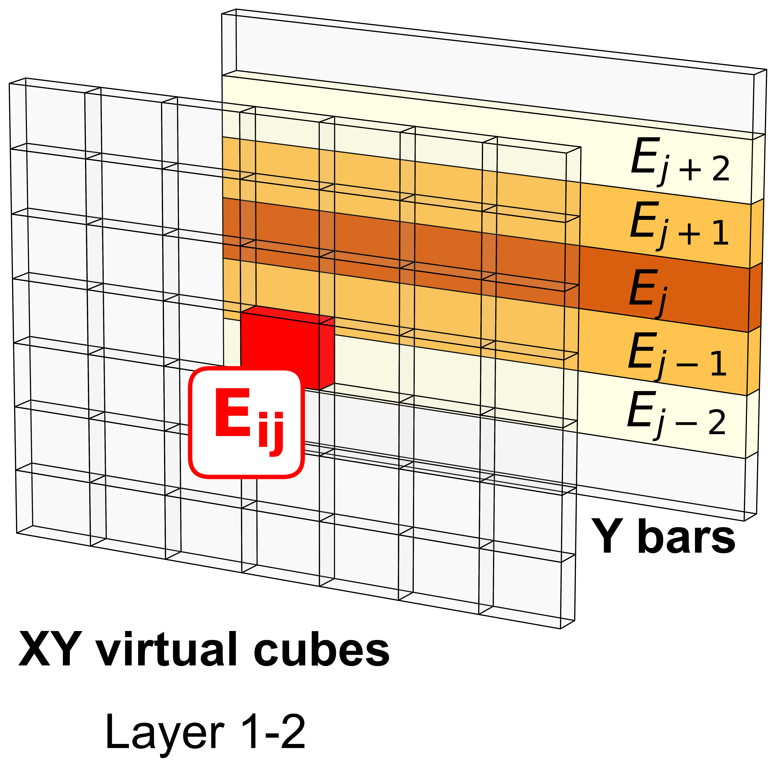}
    \caption{}
    \label{fig:segmentation_d}
\end{subfigure}
\caption{Illustration of three-dimensional energy deposition information; (a) Orthogonal X-Y bar configuration in adjacent layers. (b) An individual physical bar can be divided into a series of virtual cubes; (c) Energy deposition in a given crystal bar $E_i$ and energy deposits recorded by the orthogonal bars $E_{j-2}, E_{j-1}, \dots, E_{j+2}$; (d) Energy allocation for the virtual cube $ij$ based on the ratio of $E_j$ to the total energy deposited in the orthogonal bars.}
\label{fig:cross_location}
\end{figure}

Three-dimensional shower imaging is a core requirement of detectors designed for PFA.
A three-dimensional grid of virtual cubes are defined by the intersection points of orthogonal crystal bars in adjacnet layers.
As shown in Figure \ref{fig:segmentation_a}, layer 1 and layer 2 are composed of orthogonal X-oriented and Y-oriented bars, respectively.
An individual physical bar provides only two spatial coordinates but can be divided into a series of virtual cubes.
This division is determined by the segmentation of the orthogonal bars in adjacent layer, which supplies the missing coordinate in the complementary dimension. 
For example, the $i$-th X-oriented bar in Figure \ref{fig:segmentation_a} yields the $x$ and $z$-coordinates ($x_i$, $z_i$), and the three-dimensional coordinates of red virtual cube in Figure \ref{fig:segmentation_b} are ($x_i$, $y_j$, $z_i$), where the $y$-coordinate is from the $j$-th Y-oriented bar in adjacent layer.

By exploiting the three-dimensional continuity of shower development, the energy deposition in each virtual cube is derived through the correlation of energy measurements from orthogonal crystal bars.
The width and thickness of individual crystal bars are relatively small compare to the characteristic longitudinal and lateral extents of shower development.
Hence, the energy deposition in a given crystal bar could be allocated to its associated virtual cubes in proportion to the energy deposits recorded by the orthogonal bars in the adjacent layer.
The energy deposit measured by the $i$-th X-oriented crystal bar is denoted as $E_i$, and the energies recorded by the Y-oriented bars are represented as $E_{j-2}, E_{j-1}, \dots, E_{j+2}$, respectively, as shown in Figure \ref{fig:segmentation_c}.
The energy deposited in the virtual cube $E_{ij}$ is calculated by weighting $E_i$ with the ratio of $E_j$ to the total energy deposited in all Y-oriented bars of the adjacent layer:
\begin{equation}
E_{ij}^{\text{virtual cube}} = E_i \times \frac{E_j}{\sum_{k} E_k}
\label{eq:cross_location}
\end{equation}
where the index $k$ runs over the set of Y-oriented bars (e.g., ${j-2}, {j-1}, \dots, {j+2}$) that capture the transverse spread of the shower in the adjacent layer, as shown in Figure \ref{fig:segmentation_d}.
By leveraging the correlated spatial and energy-deposition information of a shower from orthogonally arranged crystal bars in adjacent layers, the detector achieves three-dimensional shower imaging essential for the PFA.

\subsection{Ambiguity Removal}

\begin{figure}[h]
\centering
\begin{subfigure}[t]{0.35\textwidth}
\includegraphics[width=\textwidth]{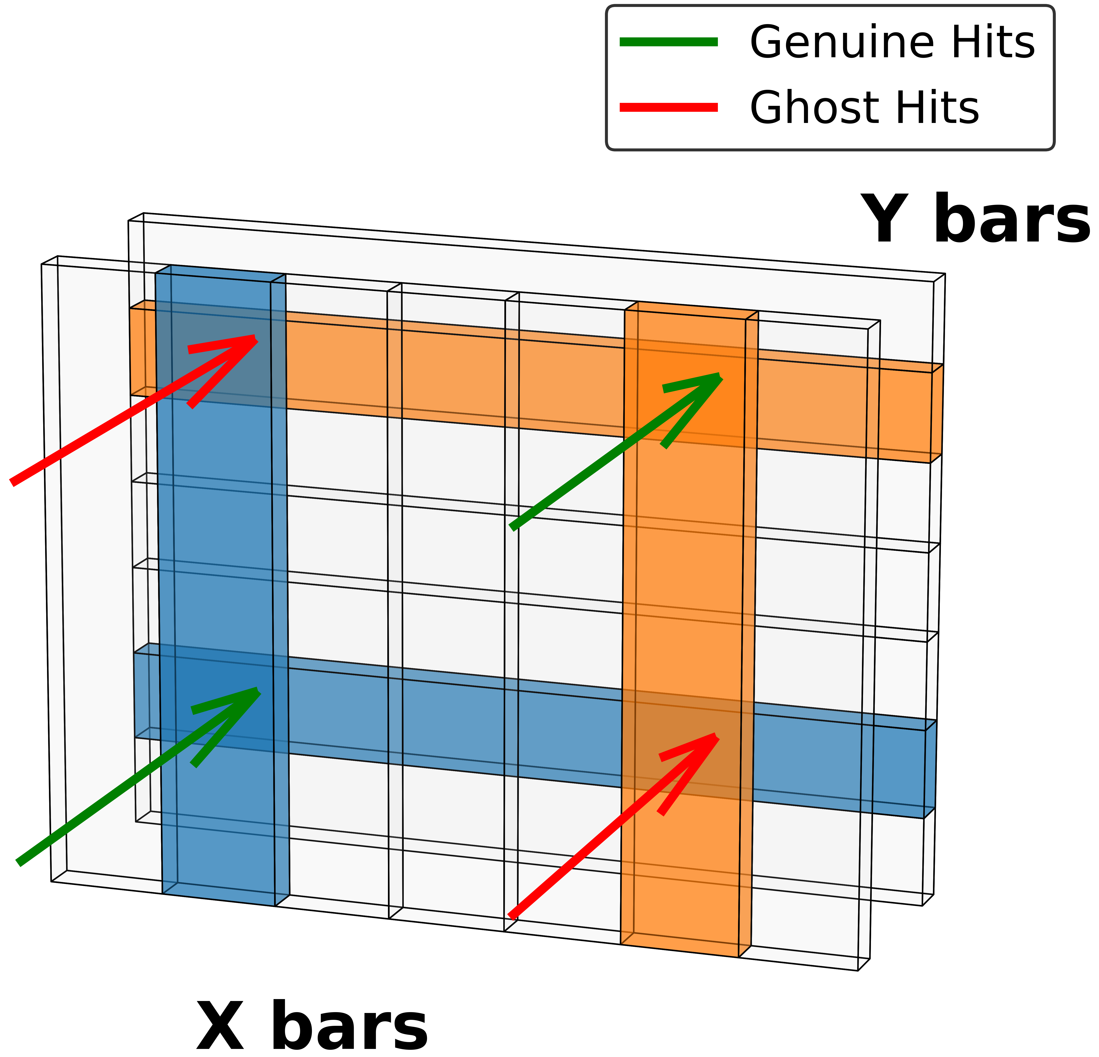}
    \caption{}
    \label{fig:ghost_a}
\end{subfigure}
\begin{subfigure}[t]{0.35\textwidth}
\includegraphics[width=\textwidth]{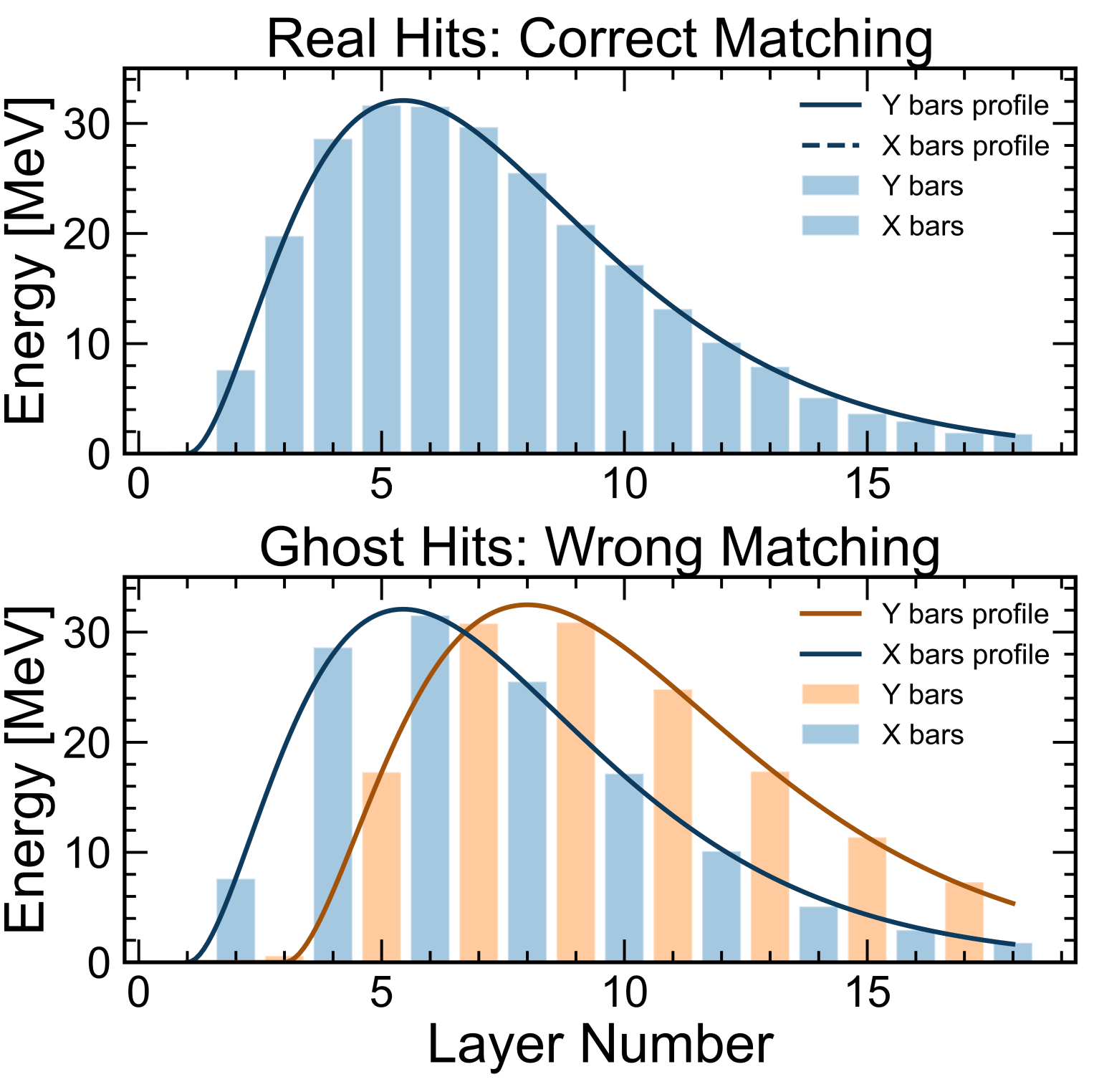}
    \caption{}
    \label{fig:ghost_b}
\end{subfigure}

\caption{Illustration of ambiguity removal. (a) Two genuine hits (green arrows) and two spurious ghost hits (red arrows) arise from the orthogonal bar geometry; (b) Comparison of longitudinal energy profiles: genuine hits show consistent profiles between X and Y bars (top), while ghost hits yield significant discrepancies (bottom).}
\label{fig:ambiguity_removal}
\end{figure}

When multiple particles are incident on a module simultaneously, correlating measurements from orthogonal crystal bars produces spurious intersections known as "ghost hits," which do not correspond to any real particle trajectory.
The continuity of the longitudinal energy profile is a defining and universal feature of shower development in matter, inherent to both electromagnetic and hadronic cascades. 
This property provides a powerful constraint for resolving ambiguities that arise from the orthogonal bar geometry.
As a representative case, Figure~\ref{fig:ghost_a} illustrates the scenario involving two incident particles in two orthogonal layers.
From the correlation of X and Y-oriented bars, a total of four intersection points are generated: two corresponding to genuine hits and two to spurious ghost hits.
Signals of same oriented crystal bars originating from a common shower in different longitudinal layers can be reliably grouped based on their spatial positions.
For a genuine trajectory, the longitudinal energy profiles of the corresponding X and Y-oriented bars exhibit strong consistency; in contrast, ambiguous combinations yield significant discrepancy between the two profiles, as depicted in Figure~\ref{fig:ghost_b}.
This ambiguity is resolved through the use of the inherent continuity of the longitudinal shower profile.
The identification of genuine hits is further refined by integrating complementary information, such as track extrapolation from the tracker and time measurements, as described in~\cite{CyberPFA}.

\subsection{Energy linearity and resolution}

\begin{figure}[h]
\centering
\begin{subfigure}[t]{0.48\textwidth}
\includegraphics[width=\textwidth]{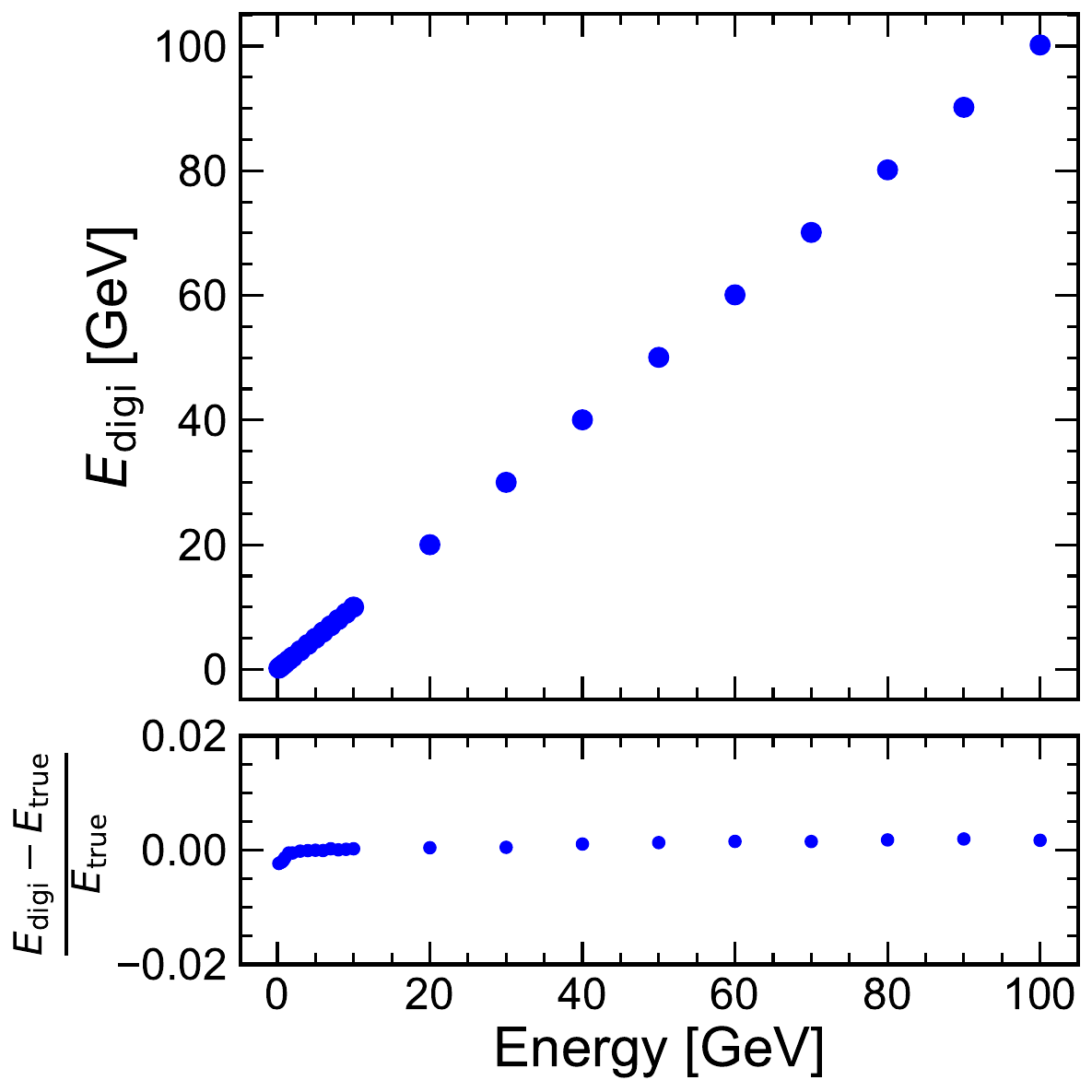}
    \caption{}
    \label{fig:EnergyScalePaper}
\end{subfigure}
\hfill
\begin{subfigure}[t]{0.48\textwidth}
\includegraphics[width=\textwidth]{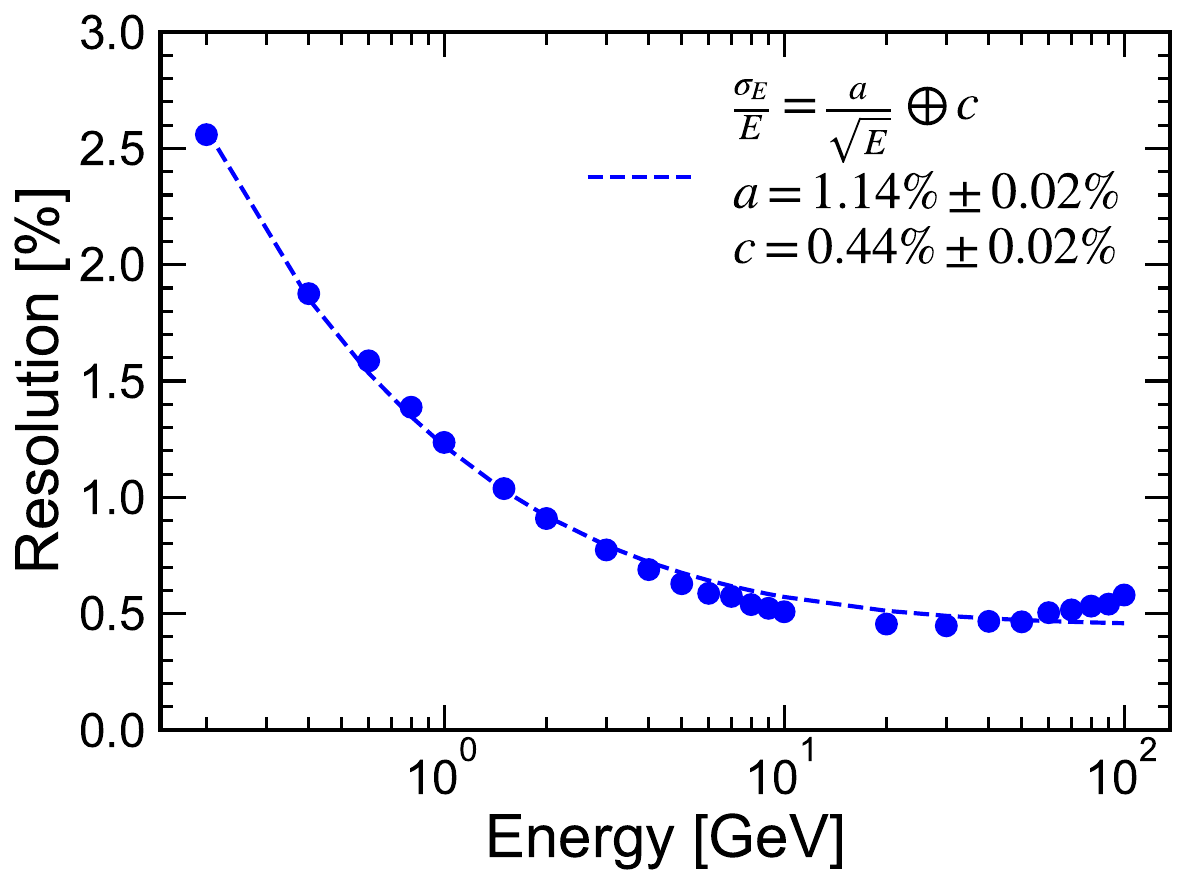}
    \caption{}
\label{fig:EnergyResolutionPaper}
\end{subfigure}
\caption{Energy linearity (a) and energy resolution (b) for simulated single-photon events.}
\label{fig:EnergyResolution}
\end{figure}

The photon energy linearity and resolution are studied using simulated single-photon events spanning an energy range from 0.2 GeV to 100 GeV.
The polar angle is restricted to $\theta \in [40^\circ, 140^\circ]$ to exclude the boundary regions, and the azimuthal angle is unrestricted over $\phi \in [0^\circ, 360^\circ]$.
The EM energy linearity is well below 1\% as shown in Figure~\ref{fig:EnergyScalePaper}. The EM energy resolution is
\begin{equation}
    \dfrac{\sigma_{E}}{E} =1.14\%/\sqrt{E} \oplus 0.44\%
    \label{eq:ECAL-Resolution}
\end{equation}
as shown in Figure~\ref{fig:EnergyResolutionPaper}.
The results of single-photon energy linearity and resolution demonstrates that the proposed ECAL design preserves the excellent intrinsic energy resolution of this crystal ECAL design.

\section{Summary}\label{summary}

A key performance metric for future $e^+e^-$ collider detectors is the jet energy resolution, which relies on three-dimensional energy deposition information for PFA reconstruction. 
The excellent intrinsic energy resolution of crystal ECALs directly enhances this metric by enabling precise measurements of photon energies within jets. 
However, conventional crystal ECALs, lacking longitudinal segmentation, cannot provide the three-dimensional spatial information essential for PFA.
We have demonstrated in this work major innovations in the simultaneous achievement of high energy resolution and three-dimensional shower imaging through a novel crystal ECAL design based on an orthogonal arrangement of crystal bars and an interleaved structure of regular and inverted trapezoidal modules.
By utilizing orthogonal crystal bars for spatial segmentation instead of fine crystal cubes, this design significantly reduces the number of electronic readout channels, thereby lowering project costs and alleviating engineering challenges related to mechanical support and cooling.
The core parameters of the barrel ECAL design, which encompass the selection of the crystal material, the crystal thickness, the transverse granularity and longitudinal segmentation, as well as the layout and modular arrangement of the crystals, have been determined to meet the stringent performance requirements of the CEPC.
Simulation of this barrel ECAL design demonstrates an EM energy resolution of $1.14\%/\sqrt{E} \oplus 0.44\%$.
The ECAL design proposed in this paper constitutes a progressive advancement in the application of crystal calorimeters for future lepton collider experiments and opens new perspectives for integrating a high-resolution crystal ECAL into the PFA paradigm.

\bibliographystyle{JHEP}
\bibliography{biblio.bib}

@article{ATLAS:2012yve,
    author = "{ATLAS Collaboration} and Aad, Georges and others",
    title = "{Observation of a new particle in the search for the Standard Model Higgs boson with the ATLAS detector at the LHC}",
    eprint = "arXiv:1207.7214",
    archivePrefix = "arXiv",
    primaryClass = "hep-ex",
    reportNumber = "CERN-PH-EP-2012-218",
    doi = "10.1016/j.physletb.2012.08.020",
    journal = "Phys. Lett. B",
    volume = "716",
    pages = "1--29",
    year = "2012"
}

@article{CMS:2012qbp,
    author = "{CMS Collaboration} and Chatrchyan, Serguei and others",
    title = "{Observation of a New Boson at a Mass of 125 GeV with the CMS Experiment at the LHC}",
    eprint = "arXiv:1207.7235",
    archivePrefix = "arXiv",
    primaryClass = "hep-ex",
    reportNumber = "CMS-HIG-12-028, CERN-PH-EP-2012-220",
    doi = "10.1016/j.physletb.2012.08.021",
    journal = "Phys. Lett. B",
    volume = "716",
    pages = "30--61",
    year = "2012"
}

@book{deBlas:2944678,
      author        = "J. de Blas et al.",
      title         = "{Physics briefing book}",
      volume        = "8",
      series        = "CERN Yellow Reports: Monograpghs",
      year          = "2025",
      url           = "https://cds.cern.ch/record/2944678",
      note          = "320 pages",
      doi           = "10.17181/CERN.35CH.2O2P",
}

@article{P5:2023wyd,
    author = "{P5 Collaboration} and Asai, Shoji and others",
    title = "{Exploring the Quantum Universe: Pathways to Innovation and Discovery in Particle Physics}",
    eprint = "arXiv:2407.19176",
    archivePrefix = "arXiv",
    primaryClass = "hep-ex",
    doi = "10.2172/2368847",
    month = "12",
    year = "2023"
}

@article{CEPCStudyGroup:2023quu,
    author = "Abdallah, Waleed and others",
    title = "{CEPC Technical Design Report: Accelerator}",
    eprint = "arXiv:2312.14363",
    archivePrefix = "arXiv",
    primaryClass = "physics.acc-ph",
    reportNumber = "IHEP-CEPC-DR-2023-01, IHEP-AC-2023-01",
    doi = "10.1007/s41605-024-00463-y",
    journal = "Radiat. Detect. Technol. Methods",
    volume = "8",
    number = "1",
    pages = "1--1105",
    year = "2024",
    note = "[Erratum: Radiat.Detect.Technol.Methods 9, 184--192 (2025)]"
}

@techreport{GuimarãesdaCosta:2678417,
      author        = "Guimarães da Costa, João Barreiro and others",
      editor        = "Guimarães da Costa, João Barreiro",
      title         = "{CEPC Conceptual Design Report: Volume 2 - Physics and Detector}",
      archivePrefix = "arXiv",
      eprint        = "1811.10545",
      reportNumber  = "IHEP-CEPC-DR-2018-02, IHEP-EP-2018-01, IHEP-TH-2018-01",
      year          = "2018",
      url           = "https://cds.cern.ch/record/2678417",
      note          = "424 pages",
}

@article{FCC:2018evy,
    author = "{FCC Collaboration} and Abada, A. and others",
    title = "{FCC-ee: The Lepton Collider}: {Future Circular Collider Conceptual Design Report Volume 2}",
    reportNumber = "CERN-ACC-2018-0057",
    doi = "10.1140/epjst/e2019-900045-4",
    journal = "Eur. Phys. J. ST",
    volume = "228",
    number = "2",
    pages = "261--623",
    year = "2019"
}

@article{Behnke:2013xla,
      title={The International Linear Collider Technical Design Report - Volume 1: Executive Summary}, 
      author={Behnke, Ties and others},
      year={2013},
      eprint={arXiv:1306.6327},
      archivePrefix={arXiv},
      primaryClass={physics.acc-ph},
      url={https://arxiv.org/abs/1306.6327}, 
}

@book{Aicheler:2012bya,
      author        = "Aicheler, M and others",
      title         = "{A Multi-TeV Linear Collider Based on CLIC Technology}",
      publisher     = "CERN",
      address       = "Geneva",
      series        = "CERN Yellow Reports: Monographs",
      year          = "2012",
      url           = "https://cds.cern.ch/record/1500095",
      doi           = "10.5170/CERN-2012-007",
}

@article{Thomson:2007zza,
    author = "Thomson, Mark Andrew",
    editor = "Albrow, M. and Raja, R.",
    title = "{Particle flow calorimetry at the ILC}",
    doi = "10.1063/1.2720472",
    journal = "AIP Conf. Proc.",
    volume = "896",
    number = "1",
    pages = "215--224",
    year = "2007"
}

@article{CALICE:2008kht,
    author = "{CALICE Collaboration} and Adloff, C. and others",
    title = "{Response of the CALICE Si-W electromagnetic calorimeter physics prototype to electrons}",
    eprint = "arXiv:0811.2354",
    archivePrefix = "arXiv",
    primaryClass = "physics.ins-det",
    reportNumber = "CALICE Analysis Note CAN-008, CALICE-CAN-2008-002, CALICE-PUB-2008-002",
    doi = "10.1016/j.nima.2009.07.026",
    journal = "Nucl. Instrum. Meth. A",
    volume = "608",
    pages = "372--383",
    year = "2009"
}

@article{SiWECAL,
    author = "{CALICE Collaboration} and Repond, J. and others",
    title = "{Design and Electronics Commissioning of the Physics Prototype of a Si-W Electromagnetic Calorimeter for the International Linear Collider}",
    eprint = "arXiv:0805.4833",
    archivePrefix = "arXiv",
    primaryClass = "physics.ins-det",
    reportNumber = "CALICE-PUB-2008-001",
    doi = "10.1088/1748-0221/3/08/P08001",
    journal = "JINST",
    volume = "3",
    pages = "P08001",
    year = "2008"
}

@article{PandoraPFA,
    author = "Thomson, M. A.",
    title = "{Particle Flow Calorimetry and the PandoraPFA Algorithm}",
    eprint = "arXiv:0907.3577",
    archivePrefix = "arXiv",
    primaryClass = "physics.ins-det",
    reportNumber = "CU-HEP-09-11",
    doi = "10.1016/j.nima.2009.09.009",
    journal = "Nucl. Instrum. Meth. A",
    volume = "611",
    pages = "25--40",
    year = "2009"
}

@techreport{CMSHGCAL,
      collaboration = "CMS",
      title         = "{The Phase-2 Upgrade of the CMS Endcap Calorimeter}",
      institution   = "CERN",
      reportNumber  = "CERN-LHCC-2017-023, CMS-TDR-019",
      address       = "Geneva",
      year          = "2017",
      url           = "https://cds.cern.ch/record/2293646",
      doi           = "10.17181/CERN.IV8M.1JY2",
}

@article{BESIII:2009fln,
    author = "{BESIII Collaboration} and Ablikim, M. and others",
    title = "{Design and Construction of the BESIII Detector}",
    eprint = "arXiv:0911.4960",
    archivePrefix = "arXiv",
    primaryClass = "physics.ins-det",
    doi = "10.1016/j.nima.2009.12.050",
    journal = "Nucl. Instrum. Meth. A",
    volume = "614",
    pages = "345--399",
    year = "2010"
}

@article{Belle-II:2010dht,
    author = "{Belle-II Collaboration} and Abe, T. and others",
    title = "{Belle II Technical Design Report}",
    eprint = "arXiv:1011.0352",
    archivePrefix = "arXiv",
    primaryClass = "physics.ins-det",
    reportNumber = "KEK-REPORT-2010-1",
    month = "11",
    year = "2010"
}

@techreport{CMS:2006myw,
  author       = {Bayatian, G. L. and others},
  title        = {CMS Physics},
  institution  = {Fermi National Accelerator Laboratory (FNAL), Batavia, IL (United States)},
  doi          = {10.2172/2510878},
  url          = {https://www.osti.gov/biblio/2510878},
  place        = {United States},
  year         = {2005},
  month        = {12}}

@article{Dong:2017/C,
  author = "Dong, Yongwei and others",
  title = "{A novel 3-D calorimeter for the High Energy cosmic-Radiation Detection (HERD) Facility onboard China’s Future Space Station}",
  doi = "10.22323/1.301.0253",
  journal = "PoS",
  year = 2017,
  volume = "ICRC2017",
  pages = "253"
}

@article{CEPCStudyGroup:2025kmw,
    author = "Adhya, Souvik Priyam and others",
    title = "{CEPC Technical Design Report - Reference Detector}",
    eprint = "arXiv:2510.05260",
    archivePrefix = "arXiv",
    primaryClass = "hep-ex",
    reportNumber = "IHEP-CEPC-DR-2025-01, IHEP-EP-2025-01",
    month = "10",
    year = "2025"
}

@article{Guo:2022wti,
    author = "Guo, Fangyi and others",
    title = "{Expected measurement precision of the branching ratio of the Higgs boson decaying to the di-photon at the CEPC}",
    eprint = "arXiv:2205.13269",
    archivePrefix = "arXiv",
    primaryClass = "hep-ex",
    reportNumber = "Chinese Physics C Vol. 47, No. 4 (2023) 043002",
    doi = "10.1088/1674-1137/acaa22",
    journal = "Chin. Phys. C",
    volume = "47",
    number = "4",
    pages = "043002",
    year = "2023"
}

@article{Ai:2024nmn,
    author = "Ai, Xiaocong and others",
    title = "{Flavor Physics at the CEPC: a General Perspective}",
    eprint = "arXiv:2412.19743",
    archivePrefix = "arXiv",
    primaryClass = "hep-ex",
    doi = "10.1088/1674-1137/adf1f0",
    journal = "Chin. Phys.",
    volume = "49",
    number = "10",
    pages = "103003",
    year = "2025"
}

@misc{CrystalCost,
    author       = {Renyuan Zhu},
    title        = {The Next Generation of
Crystal Calorimetry},
    howpublished = {The Topical Workshop on the CEPC Calorimetry},
    year         = {March 2019},
    address      = {Beijing, China},
}

@article{GEANT4:2002zbu,
    author = "Agostinelli, S. and others",
    title = "{GEANT4 - A Simulation Toolkit}",
    reportNumber = "SLAC-PUB-9350, FERMILAB-PUB-03-339, CERN-IT-2002-003",
    doi = "10.1016/S0168-9002(03)01368-8",
    journal = "Nucl. Instrum. Meth. A",
    volume = "506",
    pages = "250--303",
    year = "2003"
}

@misc{CyberPFA,
    author       = {Yang Zhang},
    title        = {CyberPFA: Particle flowalgorithmfor crystal bar ECAL},
    howpublished = {The 2025 International Workshop on the High Energy Circular Electron Positron Collider},
    year         = {November 2025},
    address      = {Guangzhou, China},
}

@article{Frank:2014zya,
    author = "Frank, Markus  and others",
    editor = "Groep, D. L. and Bonacorsi, D.",
    title = "{DD4hep: A Detector Description Toolkit for High Energy Physics Experiments}",
    doi = "10.1088/1742-6596/513/2/022010",
    journal = "J. Phys. Conf. Ser.",
    volume = "513",
    pages = "022010",
    year = "2014"
}

@article{ALLISON2016186,
    author = "Allison, J. and others",
    title = "{Recent developments in Geant4}",
    reportNumber = "FERMILAB-PUB-16-447-CD",
    doi = "10.1016/j.nima.2016.06.125",
    journal = "Nucl. Instrum. Meth. A",
    volume = "835",
    pages = "186--225",
    year = "2016"
}

@article{Qi:2025dvo,
    author = "Qi, Baohua and others",
    title = "{Development of a novel high granularity crystal electromagnetic calorimeter}",
    doi = "10.1051/epjconf/202532000012",
    journal = "EPJ Web Conf.",
    volume = "320",
    pages = "00012",
    year = "2025"
}






\end{document}